\documentclass[
	aps,
	reprint,
	superscriptaddress,
	floatfix,
	nofootinbib,
]{revtex4-2}

\usepackage{amsmath,amssymb,amsfonts,physics,graphicx,float}
\graphicspath{{pic/}}
\usepackage[setpagesize=false]{hyperref}
\allowdisplaybreaks[4]

\usepackage[labelformat=simple]{subcaption} 

\usepackage{ulem}
\usepackage{color}

\definecolor{ar}{rgb}{1.0, 0.01, 0.24}
\definecolor{al}{rgb}{0.82, 0.1, 0.26}
\definecolor{ev}{rgb}{0.56, 0.0, 1.0}

\newcommand{\lag}{\mathcal{L}}
\newcommand{\PDM}{\mathrm{PDM}}
\newcommand{\CSC}{\mathrm{CSC}}
\newcommand{\Hadron}{\mathrm{H}}
\newcommand{\Interp}{\mathrm{I}}
\newcommand{\Quark}{\mathrm{Q}}
\newcommand{\rmd}{ \mathrm{d} }

\begin{document}


\title{
Quark-hadron crossover equations of state for neutron stars: \\
constraining the chiral invariant mass in a parity doublet model 
}

\author{Takuya Minamikawa}
\email{minamikawa@hken.phys.nagoya-u.ac.jp}
\affiliation{Department of Physics, Nagoya University, Nagoya 464-8602, Japan}

\author{Toru Kojo}
\email{torujj@mail.ccnu.edu.cn}
\affiliation{Key Laboratory of Quark and Lepton Physics (MOE) and Institute of Particle Physics,
Central China Normal University, Wuhan 430079, China}

\author{Masayasu Harada}
\email{harada@hken.phys.nagoya-u.ac.jp}
\affiliation{Department of Physics, Nagoya University, Nagoya 464-8602, Japan}

\date{\today}

\begin{abstract}
We construct an equation of state (EOS) for neutron stars 
by interpolating hadronic EOS at low density 
and quark EOS at high density.
A hadronic model based on the parity doublet structure is used for hadronic matter and
a quark model of Nambu--Jona-Lasinio type is for quark matter.
We assume crossover between hadronic matter and quark matter in the the color-flavor locked phase. 
The nucleon mass of the parity doublet model has a mass associated with the chiral symmetry breaking,
and a chiral invariant mass $m_0$ which is insensitive to the chiral condensate.
The value of $m_0$ affects the nuclear EOSs at low density, and has strong correlations with the radii of neutron stars. 
Using the constraint to the radius obtained by LIGO-Virgo and NICER, 
we find that $m_0$
is restricted as 
$600\,\mathrm{MeV}\lesssim m_0 \lesssim 900\,\mathrm{MeV}$.
\end{abstract}

\maketitle


\section{Introduction}

Chiral symmetry and its spontaneous breaking is one of the most important properties 
in low-energy hadron physics.
The spontaneous breaking is triggered by the condensate of quarks and anti-quarks, 
which generates a part of hadron masses and mass difference between chiral partners.

In case of nucleon, Ref.~\cite{Detar:1988kn} introduced a notion of the 
chiral invariant mass in addition to the mass 
from the spontaneous chiral symmetry breaking using a model based on the parity doublet structure.
By regarding $N^*(1535)$ as the chiral partner to the ordinary nucleon and using the decay width, 
the chiral invariant mass is shown to be smaller than $500\,\mathrm{MeV}$~\cite{Jido:2001nt}.
On the other hand, 
analysis of nucleon mass at high temperature by lattice simulation~\cite{Aarts:2017rrl} 
suggests a large value of the chiral invariant mass.

There are many works to construct nuclear matter and neutron star (NS) EOS 
using hadronic models based on the parity doublet structure 
(see, e.g., Refs.~\cite{Hatsuda:1988mv, Zschiesche:2006zj, Dexheimer:2007tn, Dexheimer:2008cv, Sasaki:2010bp, Sasaki:2011ff,%
Gallas:2011qp, Paeng:2011hy,%
Steinheimer:2011ea,Dexheimer:2012eu, Paeng:2013xya,Benic:2015pia,Motohiro:2015taa,%
Mukherjee:2016nhb,Suenaga:2017wbb,Takeda:2017mrm,Mukherjee:2017jzi,Paeng:2017qvp,%
Marczenko:2017huu,Abuki:2018ijb,Marczenko:2018jui,Marczenko:2019trv,Yamazaki:2019tuo,Harada:2019oaq,Marczenko:2020jma,Harada:2020etl}).
Typical models are $\sigma$-$\omega$ type mean field models~\cite{Walecka:1974qa}
in which a nucleon acquires the mass from the $\sigma$ condensate,
while in parity doublet models (PDMs) nucleons are less sensitive to the details of $\sigma$ due to the presence of the chiral invariant mass. 

There have been several refinements in the PDM to account for the nucleon as well as nuclear matter properties.
The authors in Refs.~\cite{Motohiro:2015taa} and \cite{Yamazaki:2018stk} revisited the estimate of the decay width, and found that inclusion of the derivative interactions, not included in Ref. \cite{Jido:2001nt}, allows larger values of $m_0$, and discussed that relatively large values, $500\,\mathrm{MeV}\leq m_0 \leq 900\,\mathrm{MeV}$, are more reasonable to explain the saturation properties in nuclear matter.
In particular, Ref.~\cite{Motohiro:2015taa} showed that inclusion of a $\sigma^6$ term 
reproduces the incompressibility of the empirical value $K\approx240\,\mathrm{MeV}$, 
which was much larger in previous analyses. 
In Ref.~\cite{Yamazaki:2019tuo}, the analyses were further extended to NS matter, and the chiral invariant mass is restricted to be $m_0\gtrsim 600\,\mathrm{MeV}$ by the tidal deformability estimated from the NS merger GW170817~\cite{TheLIGOScientific:2017qsa,GBM:2017lvd,Abbott:2018exr}. 

The previous study in Ref.~\cite{Yamazaki:2019tuo} based on the PDM extrapolates
the hadronic equations state to the baryon density $n_B\approx3n_0$ ($n_0 \simeq 0.16\,{\rm fm}^{-3}$: nuclear saturation density). 
However, as emphasized in Refs.~\cite{Masuda:2012kf,Masuda:2012ed,Baym:2017whm,Baym:2019iky},
the validity of pure hadronic descriptions at $n_B \gtrsim 2n_0$ are questionable as nuclear many-body forces are very important, and this would imply that we need quark descriptions even before the quark matter formation.
In this context it was proposed to construct EOS by interpolating EOS for hadronic matter at $n_B\lesssim 2n_0$ and the one for quark matter in the high-density region, $n_B\gtrsim 5n_0$.
For describing the quark matter, the authors adopted a three flavor Nambu--Jona-Lasinio (NJL)-type model which leads to the color-flavor locked (CFL) color-superconducting matter, and examined effective interactions to satisfy the two-solar-mass ($2M_\odot$) constraint.
The hadronic EOSs were based on non-relativistic nuclear many-body calculations.
In Refs.~\cite{Marczenko:2019trv,Marczenko:2020jma}, 
they construct an effective model combining a PDM and
an NJL-type model with two flavors assuming no color-superconductivity.

In the present analysis, we construct EOS for NSs by interpolating the EOS constructed from the PDM proposed in Ref.~\cite{Motohiro:2015taa}, 
and the one from the NJL-type model in Refs.~\cite{Baym:2017whm,Baym:2019iky}.
Through such a construction, we will examine the properties of the PDM, especially the chiral invariant mass.
Although the nuclear and quark EOS cover different density domains, 
in fact they constrain each other as the interpolation of these EOS must satisfy the thermodynamic stability and causality constraints. 
Our unified EOSs are subject to the following NS constraints: 
the radius constraint obtained from the NS merger GW170817~\cite{TheLIGOScientific:2017qsa,GBM:2017lvd,Abbott:2018exr},
the millisecond pulsar PSR J0030+0451~\cite{Miller:2019cac,Riley:2019yda},
and the maximum mass constraint obtained from the millisecond pulsar PSR J0740+6620~\cite{Cromartie:2019kug}.

In present analyses, the most notable correlations are found between the chiral invariant mass and the radius constraints.
In the PDM, for a given $m_0$ we arranged the rest of parameters to fit the nuclear saturation properties, but the density dependence of different sets of parameters can be very different.
In particular the choice of $m_0$ affects the balance between the attractive $\sigma$ and repulsive $\omega$ interactions with nucleons; with smaller $m_0$, we need a larger scalar coupling to account for the nucleon mass, while it in turn demands a larger $\omega$ coupling for the saturation properties. As the density increases with the chiral restoration, the $\omega$ contributions become dominant, and EOSs for $n_0 \lesssim n_B \lesssim 3n_0$ become stiffer. Too stiff low density EOSs lead to too large NS radii that would contradict with the currently available upperbound. Based on this observation we will find the lowerbound for $m_0$. Meanwhile too large $m_0$ is not allowed by the nucleon mass, the lowerbound of NS radii, and the $2M_\odot$ constraint.

This paper is organized as follows:
In section~\ref{sec:formulation}, 
we explain the formulation of the present analysis.
Main results of the analysis are shown in section~\ref{sec:result}.
In section~\ref{sec:summary},
we show a summary and discussions.


\section{Formulation}
\label{sec:formulation}

In this section, 
we explain our model to determine the EOS for NSs.
In the low-density region, we use 
the parity doublet model to describe the hadronic matter.
We use the hidden local symmetry (HLS)~\cite{Bando:1987br,Harada:2003jx}
to introduce massive vector mesons with chiral symmetry. 
There are some equivalent method to the HLS~\cite{Harada:2003jx}. 

In the high-density region, on the other hand, we follow 
Refs.~\cite{Baym:2019iky,Baym:2017whm} and an NJL-type model 
with additional vector and diquark pairing interactions.
We interpolate the resultant hadronic and quark matter EOS assuming a smooth transition between them.

\subsection{Parity Doublet Model}

Here we briefly review an effective hadronic model based on the parity doublet structure for nucleons~\cite{Detar:1988kn,Jido:2001nt,Motohiro:2015taa}. 

In our model the excited nucleon $N^\ast(1535)$ is regarded as a chiral partner to the ordinary nucleon $N(939)$.
For expressing these nucleons, we introduce two baryon fields $\psi_1$ and $\psi_2$ which transform under the chiral symmetry as
\begin{align}
	\begin{aligned}
	\psi_{1}^{L} \to g_L \, \psi_1^L \ , \quad \psi_1^R \to g_R \,\psi_1^R \ , \\
	\psi_{2}^{L} \to g_R \, \psi_2^L \ , \quad \psi_2^R \to g_L \,\psi_2^R \ ,
	\end{aligned}
\end{align}
where $g_{L}$ and $g_R$ are the elements of SU(2)$_L$ and SU(2)$_R$ groups, respectively.
Two baryon fields $\psi_i^{L,R}$ ($i=1,2$) are defined as 
\begin{align}
	\psi_i^L=\frac12(1-\gamma_5)\psi_i \ , \quad 
	\psi_i^R=\frac12(1+\gamma_5)\psi_i \ .
\end{align}
We assign positive parity for $\psi_1$ and negative parity for $\psi_2$:
\begin{align}
\psi_1 \ \mathop{\rightarrow}_{P} \ \gamma_0 \psi_1 \ , \quad \psi_2 \ \mathop{\rightarrow}_{P} \ - \gamma_0 \psi_2 \ .
\end{align}
The iso-singlet scalar meson $\sigma$ and the iso-triplet pions are introduced through a $2\times2$ matrix field $M$ which transforms as
\begin{align}
M \to g_L \, M \, g_R^\dag\ .
\end{align}
In the present analysis, following Ref.~\cite{Motohiro:2015taa}, we include vector mesons based on the framework of the HLS, by decomposing the $M$ field as
\begin{align}
	M=\xi_L^\dag \, \sigma \, \xi_R \ , 
\end{align}
where $\sigma$ is the iso-singlet scalar meson field (not a matrix), 
$\xi_{L,R}$ are matrix fields including pions.
The $\xi_{L,R}$ transform under the chiral symmetry and the HLS as
\begin{align}
\xi_L \to h \, \xi_L \, g_L^\dag \ , \quad \xi_R \to h \, \xi_R \, g_R^\dag \ ,
\end{align}  
where $h$ is an element of the U(2) group for the HLS.  
In the unitary gauge of the HLS, the $\xi_{L,R}$ are parametrized as
\begin{align}
\xi_L = e^{- i \pi/f_\pi} \ , \quad \xi_R = e^{i\pi/f_\pi} \ ,
\end{align}
where $\pi$ is a $2\times2$ matrix field for pions expressed as $\pi = \sum_{a=1,2,3} \,\pi^a T_a$ with $T_a=\tau_a/2$ being the SU(2) generators and $\tau_a$ being the Pauli matrices.
For constructing the Lagrangian, it is convenient to introduce the 1-forms as 

\begin{align}\label{CMcov}
	\begin{aligned}
	\hat\alpha_\mu^\parallel
	&=\frac1{2i}\qty[(D_\mu\xi_R)\xi_R^\dag+(D_\mu\xi_L)\xi_L^\dag] \ , \\
	\hat\alpha_\mu^\perp
	&=\frac1{2i}\qty[(D_\mu\xi_R)\xi_R^\dag-(D_\mu\xi_L)\xi_L^\dag]\ .
	\end{aligned}
\end{align}
In the above expression, the covariant derivatives are defined as
\begin{align}
	D_\mu\xi_{L,R}&=(\partial_\mu-ig_\omega\omega_\mu T_0 -ig_\rho\rho_\mu^aT_a)\xi_{L,R}-i\xi_{L,R}\tilde{V}_\mu \ ,
\end{align}
where $T_0=1/2$ and $T_a=\tau_a/2$ are the U(2) generators, 
$\omega_\mu$ and $\rho_\mu^a$ are the gauge fields for U(1) and SU(2) HLS, 
$g_\omega$ and $g_\rho$ their gauge coupling constants.
As usual, external gauge fields for the 
chiral symmetry, $\tilde{V}_\mu$, is
introduced to keep track the correspondence between the generating functional of QCD and its effective Lagrangian of hadronic fields. After using the correspondence to constrain the form of the effective Lagrangian, we set the values of the external fields as 
\begin{align}
	\tilde{V}_\mu=
	\frac{1}{2} \begin{pmatrix} \mu_Q & 0 \\ 0 & - \mu_Q \end{pmatrix} \delta_\mu^0 \ .
\end{align}

Our effective Lagrangian for hadrons consists of a nucleon part and a meson part,
\begin{align}\label{L: PDM}
	\lag_\PDM&=\lag_N+\lag_M \ .
\end{align}
The nucleon part is given by
\begin{align}
	\lag_N
	= &\sum_{i=1,2}\,\bar\psi_ii\gamma^\mu D_\mu\psi_i \notag \\
& {} -g_1\qty(\bar\psi_1^L M\psi_1^R+\bar\psi_1^RM^\dag\psi_1^L) \notag \\  
& {} -g_2\qty(\bar\psi_2^L M^\dag\psi_2^R+\bar\psi_2^RM\psi_2^L)  \notag \\
	&-m_0(\bar\psi_1^L\psi_2^R-\bar\psi_1^R\psi_2^L-\bar\psi_2^L\psi_1^R+\bar\psi_2^R\psi_1^L) \notag \\
&+a_{V NN}\qty[\bar\psi_1^L\xi_L^\dag\gamma^\mu\hat\alpha_\mu^\parallel\xi_L\psi_1^L
+\bar\psi_1^R\xi_R^\dag\gamma^\mu\hat\alpha_\mu^\parallel\xi_R\psi_1^R] \notag \\
&+a_{V NN}\qty[\bar\psi_2^L\xi_R^\dag\gamma^\mu\hat\alpha_\mu^\parallel\xi_R\psi_2^L
+\bar\psi_2^R\xi_L^\dag\gamma^\mu\hat\alpha_\mu^\parallel\xi_L\psi_2^R] \notag \\
&+a_{0NN}\sum_{i=1,2}\qty[\bar\psi_i^L\gamma^\mu\tr(\hat\alpha_\mu^\parallel)\psi_i^L
	+\bar\psi_i^R\gamma^\mu\tr(\hat\alpha_\mu^\parallel)\psi_i^R] \ ,
\end{align}
where the covariant derivatives on the nucleon fields are defined as
\begin{align}
	D_\mu\psi^{L,R}_{1,2}&=(\partial_\mu-i V_\mu)\psi^{L,R}_{1,2} \,,
\end{align}
with
\begin{align}
	V_\mu =
	\begin{pmatrix} \mu_B + \mu_Q & 0 \\ 0 & \mu_B \end{pmatrix} \delta_\mu^0 \ .
\end{align}
The meson part is given by
\begin{align}
	\lag_M=\lag_M^\mathrm{kin}-V_M-V_\mathrm{SB}
	+\lag_M^\mathrm{vector}\ , 
\end{align}
where 
$\lag_M^\mathrm{kin}$, $V_M$ and $V_\mathrm{SB}$, 
are the kinetic term, the chiral symmetric potential and the potential including the explicit chiral symmetry breaking for the scalar and pseudo-scalar mesons, respectively, and 
$\lag_M^\mathrm{vector}$ includes the kinetic and mass terms for vector mesons.
The kinetic and potential terms for the scalar and pseudo-scalar mesons are expressed as~\cite{Motohiro:2015taa}
\begin{align}
	\lag_M^\mathrm{kin}=&\frac14\tr[D_\mu MD^\mu M^\dag]
	=\frac12\partial_\mu\sigma\partial^\mu\sigma
	+\sigma^2\tr[\hat\alpha_\mu^\perp\hat\alpha^\mu_\perp] \ , \\
	V_M=&-\frac14\bar\mu^2\tr[MM^\dag]+\frac1{16}\lambda_4(\tr[MM^\dag])^2 \nonumber \\
	& {} 	-\lambda_6\frac1{48}(\tr[MM^\dag])^3 \ , \\
	V_\mathrm{SB}= & -\frac14m_\pi^2f_\pi\tr[M+M^\dag] \ .
\end{align}
The vector mesons part $\lag_M^\mathrm{vector}$ is given by
\begin{align}
	\lag_M^\mathrm{vector}
	=& -\frac{1}{4}\omega_{\mu\nu}\omega^{\mu\nu}
	+\frac{m_\omega^2}{2g_\omega^2}\tr[\hat\alpha^\parallel_\mu]\tr[\hat\alpha_\parallel^\mu] \nonumber \\
	& {}-\frac1{2}\tr[\rho_{\mu\nu}\rho^{\mu\nu}] \nonumber \\
	& {} +\frac{m_\rho^2}{g_\rho^2}\qty(\tr[\hat\alpha^\parallel_\mu\hat\alpha_\parallel^\mu]
	-\frac{1}{2}
	\tr[\hat\alpha^\parallel_\mu]\tr[\hat\alpha_\parallel^\mu]) \ , 
\end{align}
where $m_\omega$ and $m_\rho$ are the masses of $\omega$ and $\rho$ mesons, 
and $\omega_{\mu\nu}$ and $\rho_{\mu\nu}$ are the field strengths of 
$\omega^\mu$ and $\rho^\mu$ respectively.
The second and forth terms include the mass terms for $\omega^\mu$ and $\rho^\mu$ as
\begin{align}
	\begin{aligned}
	\tr[\hat\alpha^\parallel_\mu]\tr[\hat\alpha_\parallel^\mu]
	&=g_\omega^2\omega^\mu\omega_\mu \ , \\
	\tr[\hat\alpha^\parallel_\mu\hat\alpha_\parallel^\mu]
	-\frac{1}{2}
	\tr[\hat\alpha^\parallel_\mu]\tr[\hat\alpha_\parallel^\mu]
	&=\frac{1}{2}g_\rho^2\rho_\mu^a\rho^\mu_a + \cdots \ ,
	\end{aligned}
\end{align}
where ``$\cdots$'' stands for interaction terms. 

In the present analysis, we calculate the thermodynamic potential in the mean field approximation as
\begin{align}
	\expval{\sigma}&=\sigma \ , \quad 
	\expval{\omega^\mu}=\omega\delta_0^\mu \ , \quad 
	\expval{\rho^\mu}=\qty(\rho -\frac{\mu_Q}{g_\rho})T_3\delta_0^\mu \ .
\end{align}
Each mean field is assumed to be independent of the spatial coordinates. 
Mean field $\rho$ is defined in such a way that $\lag_M$ does not explicitly include $\mu_Q$. 

It is convenient to introduce the effective chemical potentials of protons and neutrons as
\begin{align}
	\begin{aligned}
	\mu_p^\ast&=\mu_Q+\mu_B-g_{\omega NN}\, \omega-\frac{1}{2}g_{\rho NN}\,\rho \ , \\
	\mu_n^\ast&=\mu_B-g_{\omega NN}\, \omega+\frac{1}{2}g_{\rho NN}\,\rho \ ,
	\end{aligned}
\end{align}
where
\begin{align}
	\begin{aligned}
	g_{\omega NN}&=(a_{VNN}+a_{0NN})g_\omega \ , \\
	g_{\rho NN}&=a_{VNN}g_\rho \ .
	\end{aligned}
\end{align}

The thermodynamic potential in the hadronic matter is calculated as~\cite{Motohiro:2015taa} 
\begin{align}
	\Omega_\PDM
	&=V(\sigma)-V(f_\pi)-\frac12m_\omega^2\omega^2-\frac12m_\rho^2\rho^2 \nonumber \\
	&-2\sum_{i=1,2}\sum_{\alpha=p,n}\int^{k_F}
	\frac{\dd[3]\mathbf{p}}{(2\pi)^3}
	(\mu_\alpha^*-E^i_\mathbf{p}) \ ,\label{PDM: grand functional}
\end{align}
where $i=1$ labels the ordinary nucleon $N(939)$ and $i=2$ the excited nucleon $N^*(1535)$,   
$E_\mathbf{p}^i=\sqrt{\mathbf{p}^2+m_i^2}$ is the energy of relevant particle with mass $m_i$ and momentum $\mathbf{p}$.
In the integration above, the integral region is restricted as $|\mathbf{p}|<k_F$ where 
$k_F=\sqrt{(\mu_\alpha^\ast)^2-m_i^2}$ is the fermi momentum for the relevant particle.
We notice that we use the so called no sea approximation, assuming that the structure of the Dirac sea remains the same for the vacuum and medium. 
$V(\sigma)$ is the potential of $\sigma$ mean field, 
\begin{align}
	V(\sigma)&=-\frac12\bar\mu^2\sigma^2+\frac14\lambda_4\sigma^4
	-\frac16\lambda_6\sigma^6-m_\pi^2f_\pi\sigma \ .
\end{align}
In Eq.~(\ref{PDM: grand functional}) we subtracted the potential in vacuum $V(f_\pi)$, with which the total potential in vacuum is zero.

The total thermodynamic potential of the hadronic matter in NSs is
obtained by including the effects of leptons as
\begin{align}
	\Omega_\Hadron=\Omega_\PDM+\sum_{l=e,\mu}\Omega_l \ ,
\end{align}
where $\Omega_l$ ($l = e,\mu$) are 
the thermodynamic potentials for leptons given by
\begin{align}\label{pot: lepton}
	\Omega_l=-2\int^{k_F}
	\frac{\dd[3]\mathbf{p}}{(2\pi)^3}
	(\mu_l-E_\mathbf{p}^l) \ .
\end{align}
Here, the mean fields are determined by the following stationary conditions:
\begin{align}\label{gap: PDM}
	0=\pdv{\Omega_\Hadron}{\sigma}\ , 
	\quad 0=\pdv{\Omega_\Hadron}{\omega}\ , 
	\quad 0=\pdv{\Omega_\Hadron}{\rho} \ .
\end{align}
In NSs, we impose the beta equilibrium and the charge neutrality condition represented as
\begin{align}
\label{beta eq.}	\mu_e=\mu_\mu=-\mu_Q \ , \\
\pdv{\Omega_\Hadron}{\mu_Q}=n_p-n_l=0 \ .
\end{align}
Finally, we obtain the pressure in the hadronic matter as
\begin{equation}\label{P_H}
P_\Hadron=-\Omega_\Hadron \ .
\end{equation}

In the present analysis, following Ref.~\cite{Motohiro:2015taa}, we determine the model parameters from the following physical inputs for fixed values of the chiral invariant mass $m_0$:
five masses of the relevant hadrons and the pion decay constant in vacuum as listed in Table~\ref{input: mass}; 
\begin{table}\centering
	\caption{  {\small Physical inputs in vacuum in unit of MeV.  }  }\label{input: mass}
	\begin{tabular}{cccccc}
		\hline\hline
		$m_\pi$ & $f_\pi$ & $m_\omega$ & $m_\rho$ & $m_+$ & $m_-$\\
		\hline
		140 & 92.4 & 783 & 776 & 939 & 1535\\
		\hline\hline
	\end{tabular}
\end{table}
saturation properties of nuclear matter at the saturation density as in Table~\ref{input: saturation}.
\begin{table}\centering
	\caption{  {\small Saturation properties used to determine the model parameters: the saturation density $n_0$, the binding energy $B_0$, the incompressibility $K_0$ and the symmetry energy $S_0$.}  }
	\label{input: saturation}
	\begin{tabular}{cccc}\hline\hline
	$n_0$ [fm$^{-3}$] & $B_0$ [MeV] & $K_0$ [MeV] & $S_0$ [MeV]\\
	\hline
	0.16 & 16 & 240 & 31\\
	\hline\hline
	\end{tabular}
\end{table}
We show the values of model parameters for several typical choices of $m_0$ in Table~\ref{output}.

As is seen from Table~\ref{output}, 
the slope parameter in this model is larger for a smaller chiral invariant mass. 
Although the higher order contributions in the expansion with respect to $x=(n_B-n_0)/3n_0$ and $\delta=2n_I/n_B$ 
become important 
in the high density region $n_B\gtrsim 2n_0$, 
the EOS from the present model is stiffer for smaller $m_0$ as we will show in the next section. 
This can be understood as follows:
The Yukawa coupling of $\sigma$ to nucleon is larger for smaller chiral invariant mass, 
which leads to stronger attractive force mediated by $\sigma$ contribution.
The $\omega$ contribution causing the repulsive force
is also larger to satisfy the saturation properties at saturation density.
This $\omega$ contribution becomes larger in the high density region, 
while the $\sigma$ contribution becomes smaller.
The resulting large repulsive force makes the EOS stiff. 

\begin{table}\centering
	\caption{ 
	{\small Values of model parameters determined for several choices of $m_0$. The values of the slope parameter is also shown as output. } }\label{output}
	\begin{tabular}{c|ccccc}
		\hline\hline
		$m_0$ [MeV] & 500 & 600 & 700 & 800 & 900 \\
		\hline
		$g_1$ & 9.02 & 8.48 & 7.81 & 6.99 & 5.96\\
		$g_2$ &15.5 & 14.9 & 14.3 & 13.4 & 12.4\\
		$\bar\mu^2/f_\pi^2$ & 22.7 & 22.4 & 19.3 & 11.9 & 1.50\\
		$\lambda_4$ & 41.9 & 40.4 & 35.5 & 23.1 & 4.43\\
		$\lambda_6f_\pi^2$ & 16.9 & 15.8 & 13.9 & 8.89 & 0.636\\
		$g_{\omega NN}$ & 11.3 & 9.13 & 7.30 & 5.66 & 3.52\\
		$g_{\rho NN}$ & 7.31 & 7.86 & 8.13 & 8.30 & 8.43\\
		\hline
		$L_0$ [MeV] & 93.76 & 86.24 & 83.04 & 81.33 & 80.08\\
		\hline\hline
	\end{tabular}
\end{table}

\subsection{Color-Superconductivity}

Following Ref.~\cite{Baym:2019iky}, 
we use an NJL-type effective model of quarks 
including the 4-Fermi interactions
which cause the spontaneous chiral symmetry breaking
and the color-superconductivity.
The Lagrangian is given by
\begin{align}\label{L_CSC}
	\lag_\CSC
	&=\lag_0
	+\lag_\sigma
	+\lag_\mathrm{d}
	+\lag_\mathrm{KMT}
	+\lag_\mathrm{vec} \ ,
\end{align}
where
\begin{align}
	\lag_0&=\bar q(i\gamma^\mu\partial_\mu-\hat m_q+\gamma_\mu \hat{A}^\mu)q \ , \\
	\lag_\sigma&=G\sum_{A=0}^8\qty[(\bar q\tau_Aq)^2+(\bar qi\gamma_5\tau_Aq)^2]\ , \\
	\lag_\mathrm{d}
	&=H\sum_{A,B=2,5,7}\left[(\bar q\tau_A\lambda_BC\bar q^t)(q^tC\tau_A\lambda_Bq)\right. \nonumber \\
	&+\left.(\bar qi\gamma_5\tau_A\lambda_BC\bar q^t)(q^tCi\gamma_5\tau_A\lambda_Bq)\right]\ , \\
	\lag_\mathrm{KMT}
	&=-K\qty[\det_f\bar q(1-\gamma_5)q+\det_f\bar q(1+\gamma_5)q]\ , \\
	\lag_\mathrm{vec}
	&=-g_V(\bar q\gamma^\mu q)(\bar q\gamma_\mu q) \ ,
\end{align}
and $\hat{A}^\mu$ is the external field. 
The chemical potentials are introduced in the same way as the hadronic case by 
\begin{align}
	\hat{A}^\mu = (\mu_q+\mu_3\lambda_3+\mu_8\lambda_8+\mu_QQ)\delta^\mu_0 \ ,
\end{align}
where $\lambda_a$ are Gell-Mann matrices in color space
and $Q=\mathrm{diag}(2/3,-1/3,-1/3)$ is a charge matrix in flavor space. 
For coupling constants $G$ and $K$, we chose the values of Hatsuda-Kunihiro parameters 
which successfully reproduce the hadron phenomenology at low energy \cite{Hatsuda:1994pi,Baym:2017whm}:
$G\Lambda^2=1.835$ and $K\Lambda^5=9.29$ with $\Lambda=631.4$ MeV.
We introduce the mean fields as
\begin{align}
\sigma_f & = \ev {\bar{q}_f q_f} \ , \quad ( f =u, d, s ) \ , \\
d_j & = \ev {q^tC\gamma_5R_jq} \ , \quad ( j = 1,2,3 ) \ , \\
n_q & = \sum_{f=u,d,s} \ev{ q_f^\dag q_f } \ ,
\end{align}
where 
$(R_1,R_2,R_3)=(\tau_7\lambda_7,\tau_5\lambda_5,\tau_2\lambda_2)$. 
Then, the thermodynamic potential is calculated as
\begin{align}
	\Omega_\CSC 
	&=\Omega_s-\Omega_s[\sigma_f=\sigma^0_f,d_j=0,\mu_q=0] \nonumber \\
	&+\Omega_c-\Omega_c[\sigma_f=\sigma^0_f,d_j=0]\ , 
\end{align}
where
\begin{align}
	\Omega_s&=-2\sum_{i=1}^{18}\int^\Lambda\frac{d^3\mathbf{p}}{(2\pi)^3}\frac{\varepsilon_i}{2}\ , \label{Omega s}\\
	\Omega_c&=\sum_{i}(2G\sigma_i^2+Hd_i^2)-4K\sigma_u\sigma_d\sigma_s-g_Vn_q^2\ .
\end{align}
In Eq.~(\ref{Omega s}), 
$\varepsilon_i$ are energy eigenvalues obtained from the following inverse propagator in Nambu-Gorkov basis
\begin{align}
	S^{-1}(k)&=\mqty(\gamma_\mu k^\mu-\hat M+\gamma^0\hat\mu & \gamma_5\sum_i\Delta_iR_i \\
	-\gamma_5\sum_i\Delta_i^\ast R_i & \gamma_\mu k^\mu-\hat M-\gamma^0\hat\mu)\ , 
\label{inverse propagator}
\end{align}
where
\begin{align}
	M_i&=m_i-4G\sigma_i+K|\epsilon_{ijk}|\sigma_j\sigma_k\ , \\
	\Delta_i&=-2Hd_i\ , \\
	\hat\mu&=\mu_q-2g_Vn_q+\mu_3\lambda_3+\mu_8\lambda_8+\mu_QQ \ .
\end{align}
$S^{-1}(k)$ in Eq.~(\ref{inverse propagator}) is $72\times72$ matrix 
in terms of the color, flavor, spin and Nambu-Gorkov basis, 
which has 72 eigenvalues.
$M_{u,d,s}$ are the constituent masses of the $u,d,s$-quarks 
and $\Delta_{1,2,3}$ are the gap energies. 
In the present parameter choice, at $n_B \gtrsim 5n_0$ they vary in the range of $M_{u,d}\approx50$-$100$ MeV, $M_s\approx$ 250-300 MeV 
and $\Delta_{1,2,3}\approx$ 200-250 MeV~\cite{Baym:2017whm}. 
Note that the matix does not depend on the spin, and that the charge conjugation invariance relates two eigenvalues.
Then, there are 18 independent eigenvalues at most.

The total thermodynamical potential is
\begin{align}
	\Omega_\Quark=\Omega_\CSC+\sum_{l=e,\mu}\Omega_l \ ,
\end{align}
where $\Omega_l$ is the thermodynamic potential for leptons given in Eq.~(\ref{pot: lepton}).
The chiral condensates $\sigma_j$ and the diquark condensates $d_i$ are determined from the gap equations, 
\begin{align}
	0=\pdv{\Omega_\Quark}{\sigma_i}=\pdv{\Omega_\Quark}{d_i} \ .
\end{align}
To determine the relevant chemical potentials other than the baryon number density,
we use the beta equilibrium condition given in Eq.~(\ref{beta eq.}), and the conditions for electromagnetic charge neutrality and color charge neutrality expressed as
\begin{align}
	n_j=-\pdv{\Omega_\Quark}{\mu_j}=0 \ ,
\end{align}
where $j=3,8,Q$. 
The baryon number density $n_B$ is three times of quark number density determined as
\begin{align}
	n_q=-\pdv{\Omega_\Quark}{\mu_q} \ , 
\end{align}
where $\mu_q$ is $1/3$ of the baryon number chemical potential.
Substituting the above conditions, we obtain the pressure of the system as
\begin{equation}\label{P_Q}
	P_\Quark=-\Omega_\Quark \ .
\end{equation}


\subsection{Interpolation of EOS}

\newcommand{\figsize}{0.497\hsize}
\newcommand{\widthsize}{\hsize}
\begin{figure*}[hbt]
	\begin{subfigure}{\figsize}
		\includegraphics[width=\widthsize]{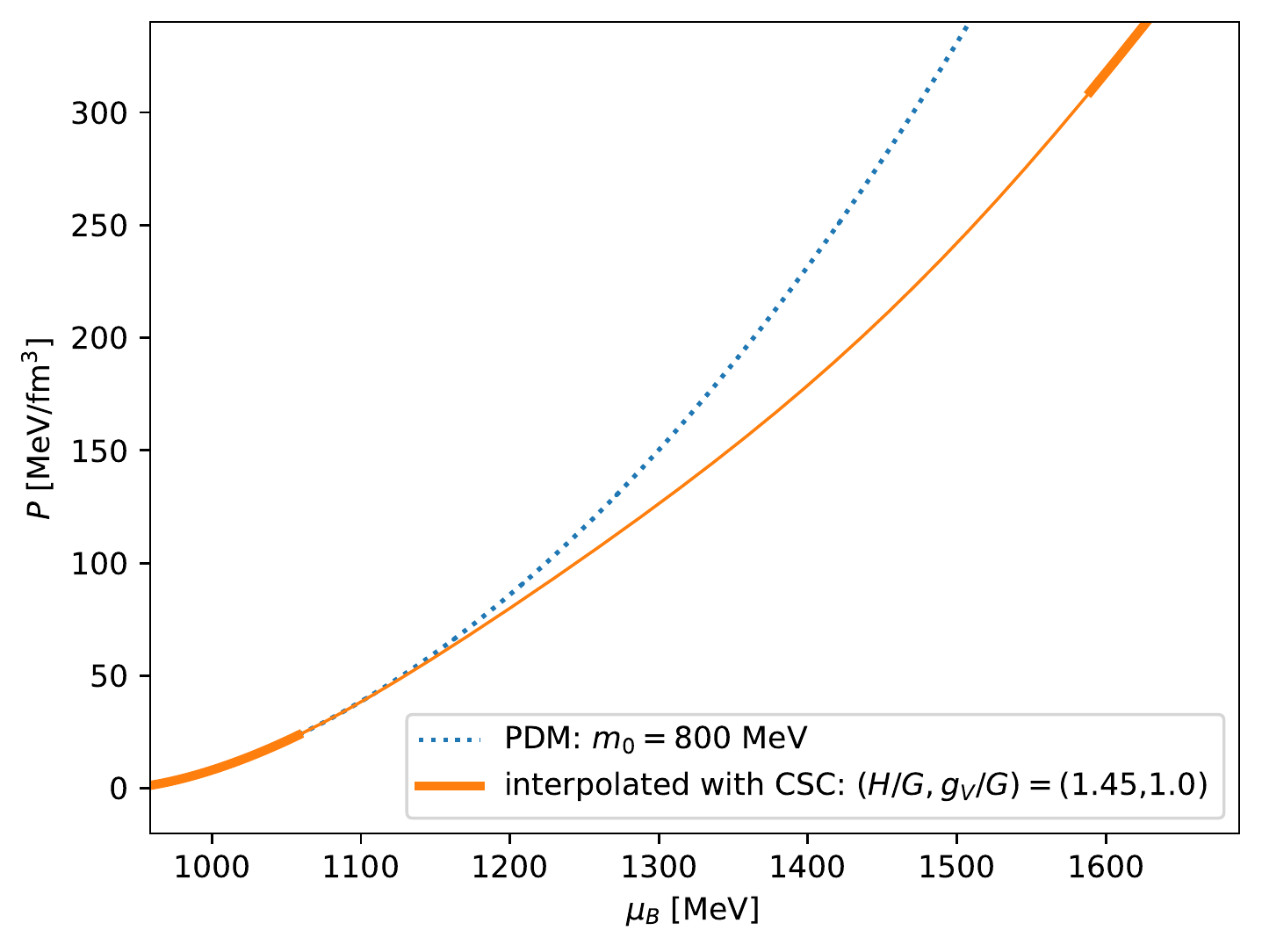}
		\caption{$(H/G,g_V/G)=(1.45,1.0)$}
		\label{P: 800-sat}
	\end{subfigure}
	\begin{subfigure}{\figsize}
		\includegraphics[width=\widthsize]{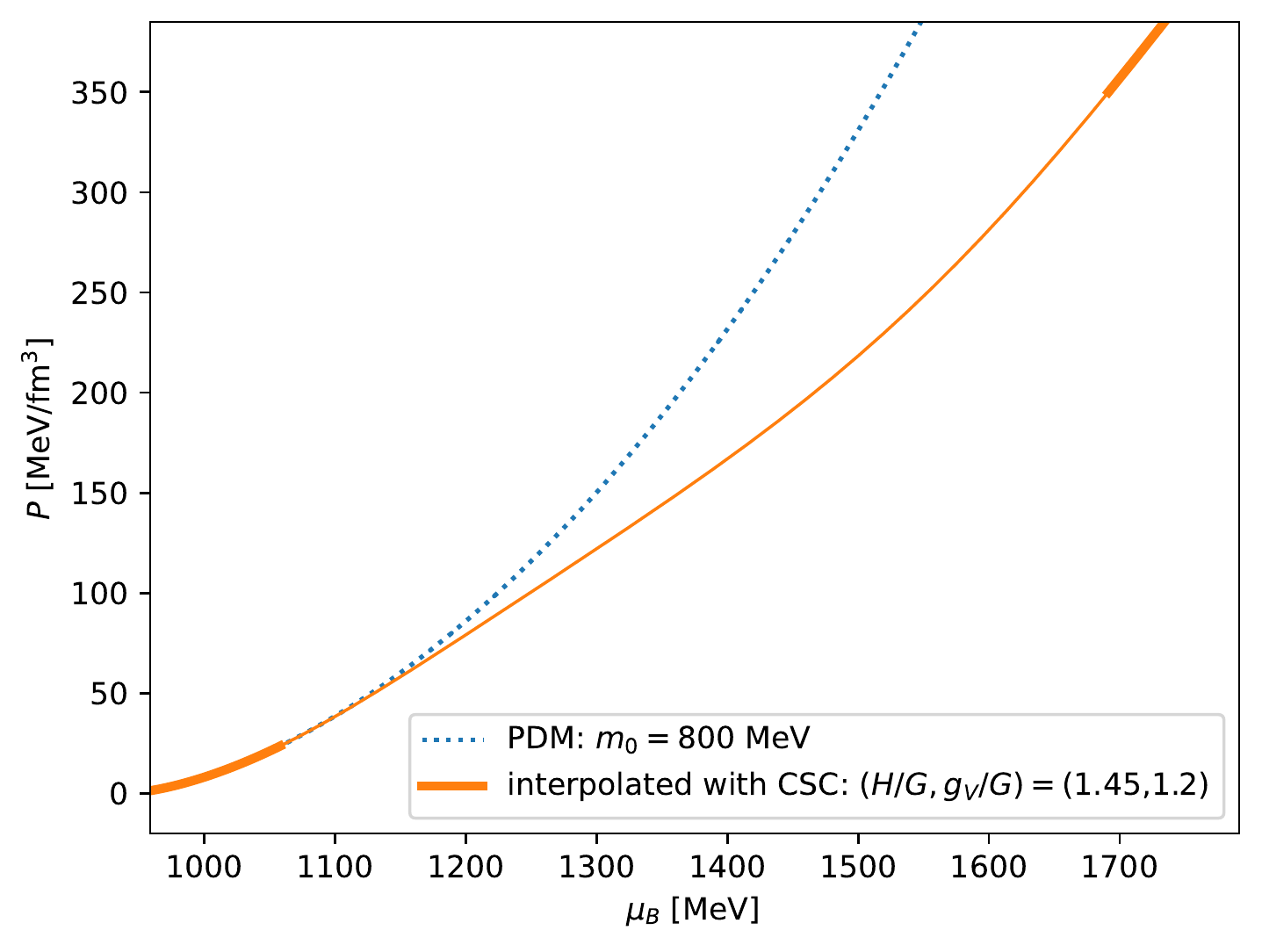}
		\caption{$(H/G,g_V/G)=(1.45,1.2)$}
		\label{P: 800-unsat}
	\end{subfigure}
\caption{  {\small 
Pressure $P(\mu_B)$ of the PDM and the unified equations of state. For the PDM we chose $m_0=800 $ MeV, and for quark models we used $(H/G,g_V/G)=(1.45,1.0)$ and $(1.45,1.2)$. The thick curves in the unified equations of state are used to mark the pure hadronic and quark parts.  
}
}
\label{P-muB}
\end{figure*}

\begin{figure*}[htb]
	\begin{subfigure}{\figsize}
		\includegraphics[width=\widthsize]{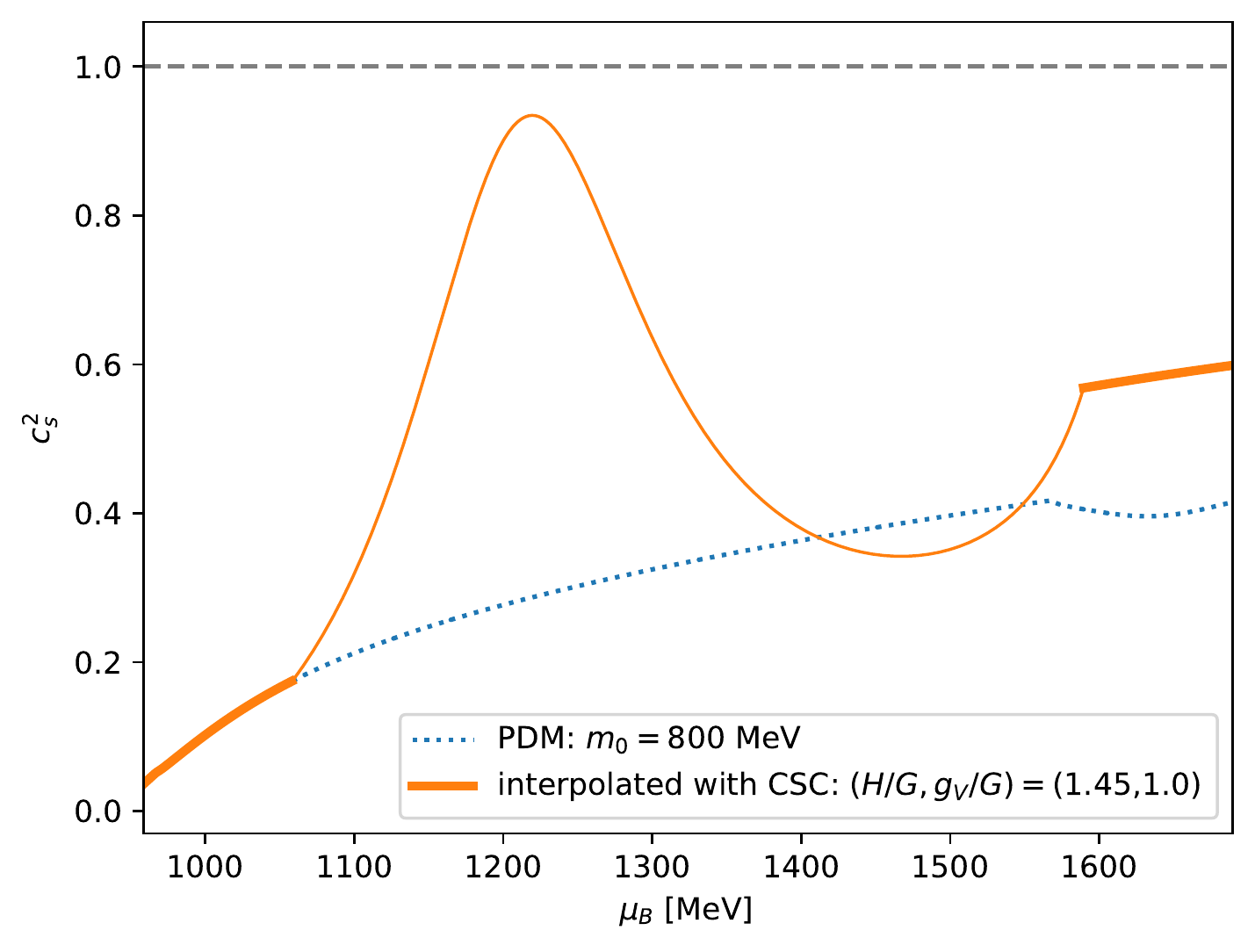}
		\caption{$(H/G,g_V/G)=(1.45,1.0)$}
		\label{cs: 800-sat}
	\end{subfigure}
	\begin{subfigure}{\figsize}
		\includegraphics[width=\widthsize]{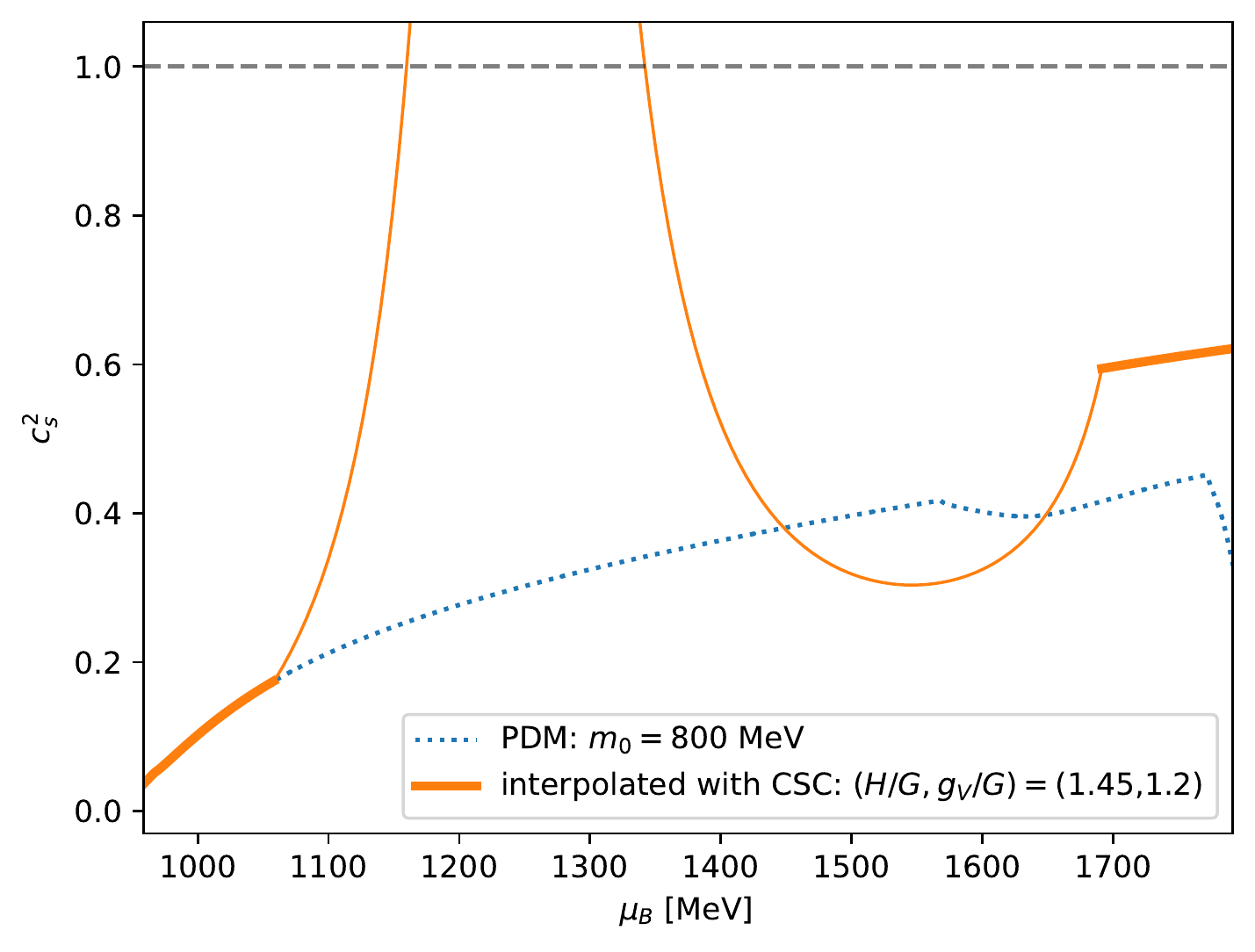}
		\caption{$(H/G,g_V/G)=(1.45,1.2)$}
		\label{cs: 800-unsat}
	\end{subfigure}
\caption{  {\small
Squared speed of sound $c_s^2$ 
for $(H/G,g_V/G)=(1.45,1.0)$ and (1.45,1.2). 
Curves are same as in Fig.~\ref{P-muB}. 
}
}
\label{cs2-muB}
\end{figure*}

\begin{figure*}[p]
	\begin{subfigure}{\figsize}
		\includegraphics[width=\widthsize]{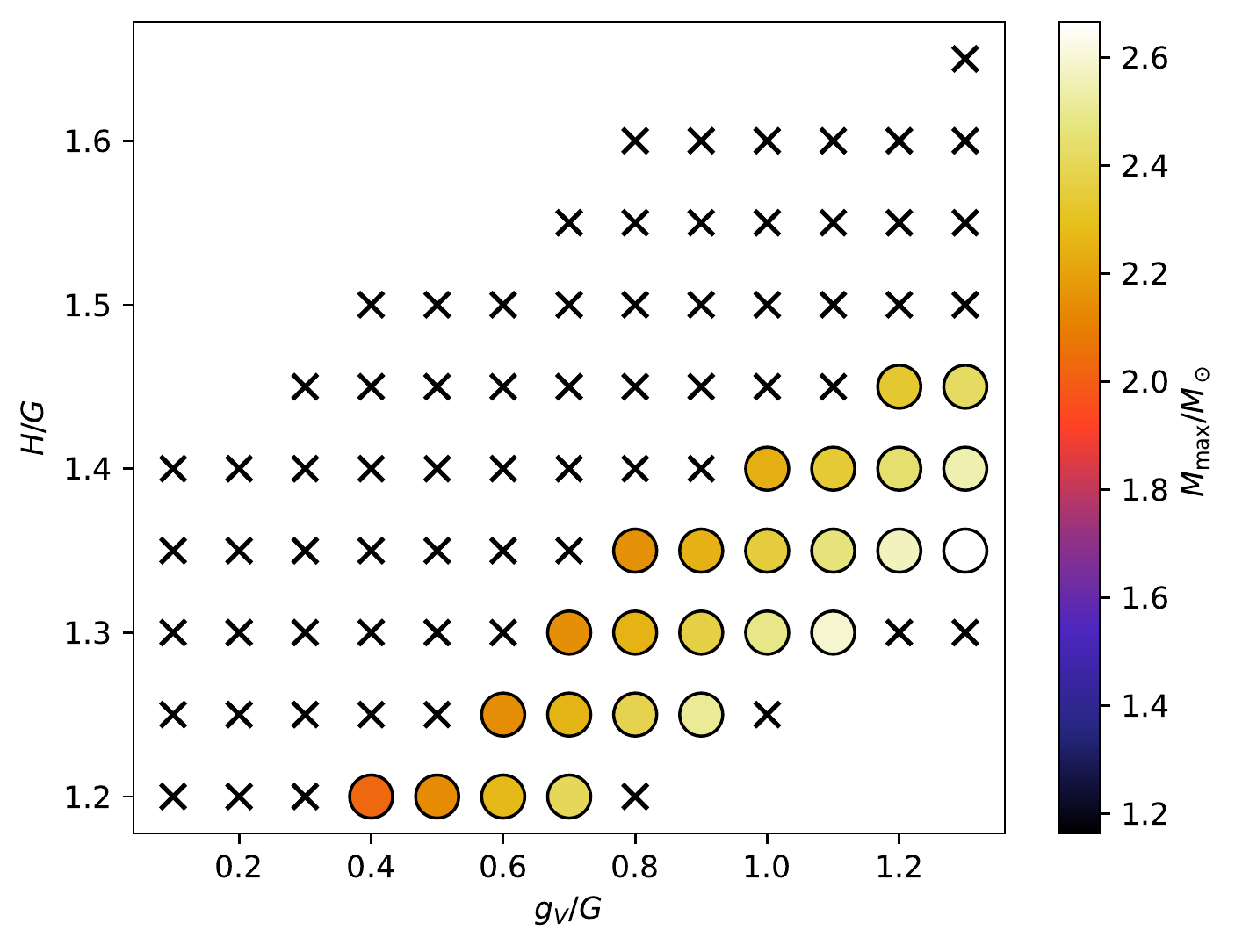}
		\caption{$m_0=500$\,MeV}
		\label{cond_check_with_Mmax_500}
	\end{subfigure}
	\begin{subfigure}{\figsize}
		\includegraphics[width=\widthsize]{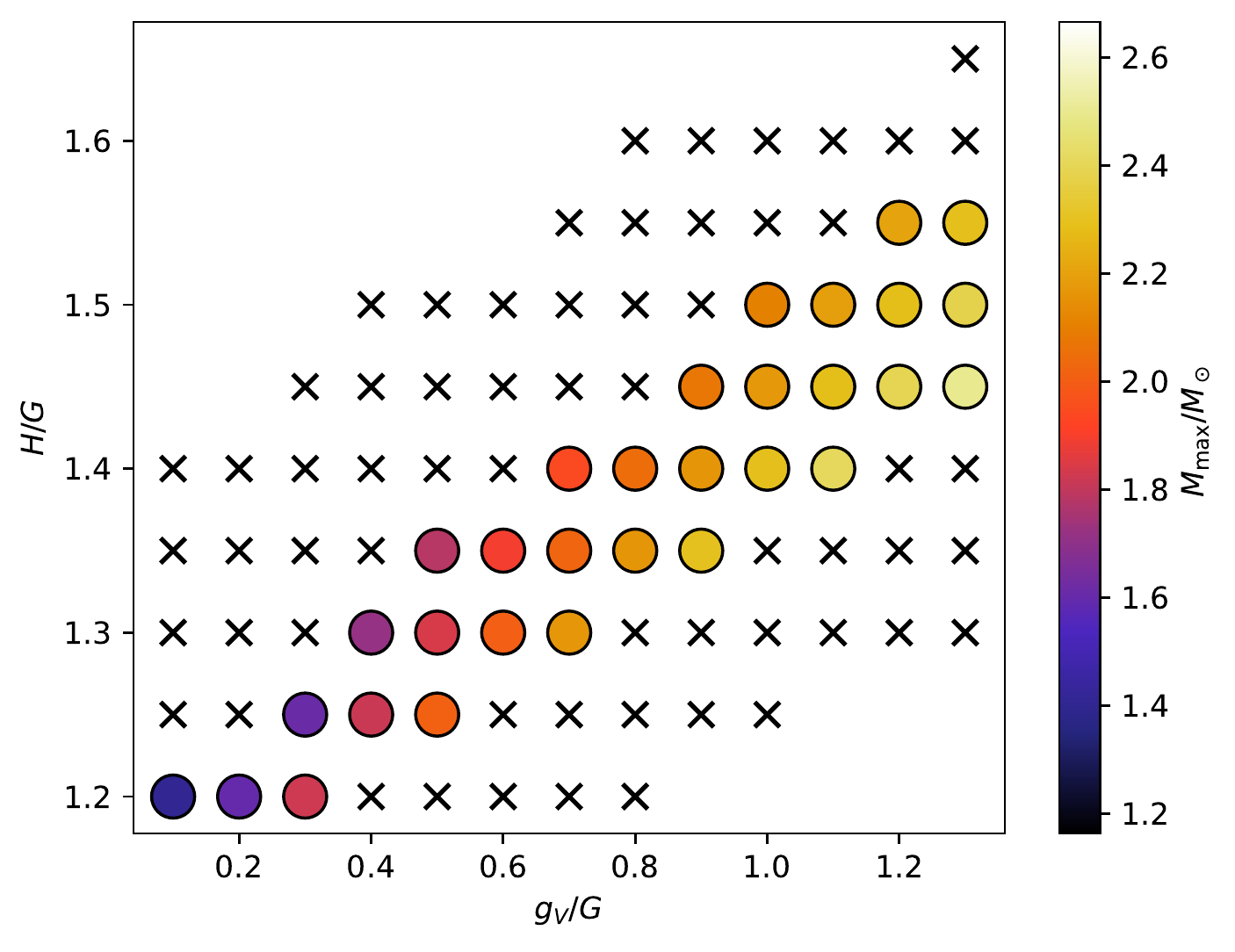}
		\caption{$m_0=600$\,MeV}
		\label{cond_check_with_Mmax_600}
	\end{subfigure}
	\begin{subfigure}{\figsize}
		\includegraphics[width=\widthsize]{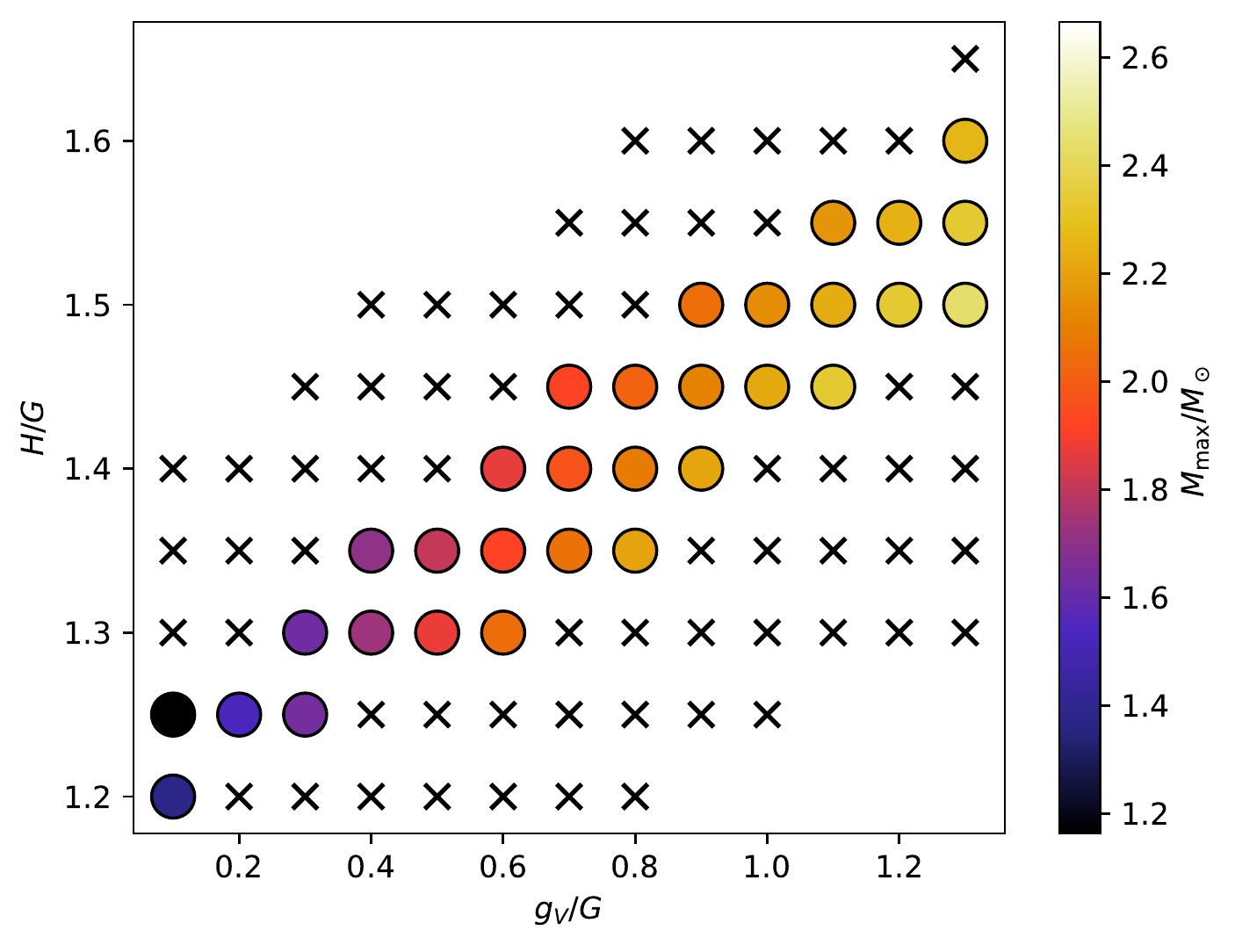}
		\caption{$m_0=700$\,MeV}
		\label{cond_check_with_Mmax_700}
	\end{subfigure}
	\begin{subfigure}{\figsize}
		\includegraphics[width=\widthsize]{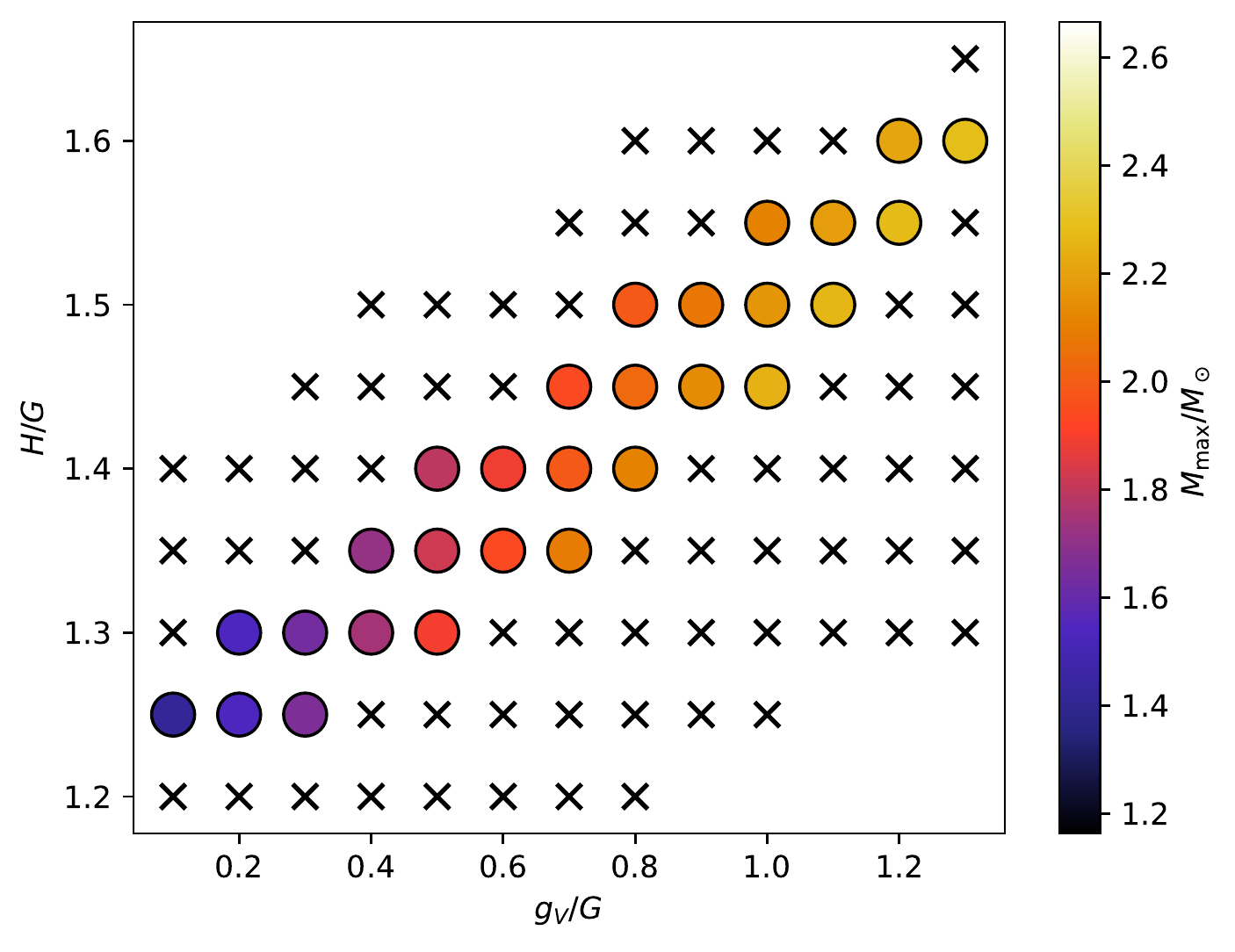}
		\caption{$m_0=800$\,MeV}
		\label{cond_check_with_Mmax_800}
	\end{subfigure}
	\begin{subfigure}{\figsize}
		\includegraphics[width=\widthsize]{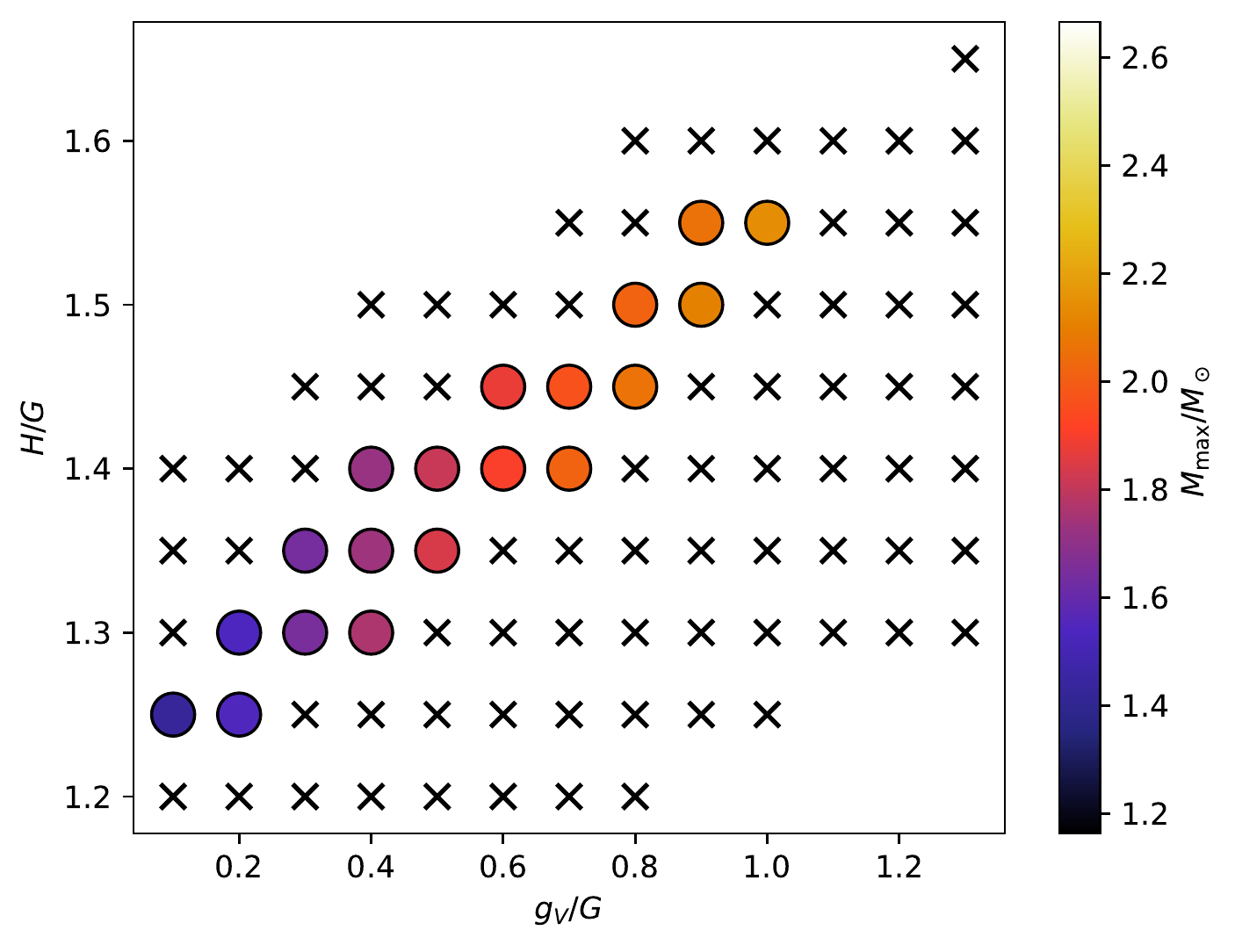}
		\caption{$m_0=900$\,MeV}
		\label{cond_check_with_Mmax_900}
	\end{subfigure}
\caption{ 
{\small
(Color online)
Allowed combinations of $(H,g_V)$ for 
\subref{cond_check_with_Mmax_500} $m_0=500$\,MeV, 
\subref{cond_check_with_Mmax_600} $m_0=600$\,MeV, 
\subref{cond_check_with_Mmax_700} $m_0=700$\,MeV, 
\subref{cond_check_with_Mmax_800} $m_0=800$\,MeV and
\subref{cond_check_with_Mmax_900} $m_0=900$\,MeV. 
Cross mark indicates that the combination of $(H,g_V)$ is excluded
by the causality constraint. 
Circle indicates that the combination is allowed. 
The color
of the circle shows the maximum mass of NS obtained
from the corresponding parameters, 
as indicated by a vertical bar at the right side of each figure. 
}
}
\label{cond_check_with_Mmax}
\end{figure*}


Here, we consider interpolation of two EOSs for 
hadronic matter and quark matter which are constructed in previous subsections.
Following Ref.~\cite{Baym:2017whm}, 
we assume that hadronic matter is realized
in the low density region
$n_B<2n_0$, 
and use the pressure constructed in Eq.~(\ref{P_H}).
In the high density region $n_B>5n_0$, 
on the other hand, 
the pressure given in Eq.~(\ref{P_Q}) of quark matter is used. 
In the intermediate region 
$2n_0<n_B<5n_0$, 
we assume that the pressure is expressed by a fifth order polynomial of $\mu_B$ as
\begin{align}
	P_\Interp(\mu_B)=\sum_{i=0}^5C_i\mu_B^i \ , 
\end{align}
where $C_i$ are six fee parameters to be determined from the following boundary conditions, 
\begin{align}
&\frac{ \rmd^n P_{ {\rm I}} }{ (\rmd \mu_B )^n } \bigg|_{ \mu_{ BL} } =\frac{ \rmd^n P_{\Hadron} }{ (\rmd \mu_B)^n } \bigg|_{ \mu_{ BL} } \,, \nonumber \\
&\frac{ \rmd^n P_{ {\rm I}} }{ (\rmd \mu_B )^n } \bigg|_{ \mu_{ BU} } =\frac{ \rmd^n P_{\Quark} }{ (\rmd \mu_B)^n } \bigg|_{ \mu_{ BU} } \,,~~~ (n=0,1,2) 
\end{align}
where $\mu_{BL}$ is the chemical potential 
corresponding to $n_B=2n_0$ 
and $\mu_{BU}$ to  $n_B=5n_0$. 

We show typical examples of the connected pressure in Fig.~\ref{P-muB}
and corresponding sound velocity calculated by
\begin{align}
	c_s^2=\dv{P}{\varepsilon}=\frac{n_B}{\mu_B\chi_B} \ ,
\end{align}
where 
$n_B=\dv{P}{\mu_B}$ and 
$\chi_B=\dv[2]{P}{\mu_B}$
in Fig.~\ref{cs2-muB}.
We see that, although both plots \ref{P: 800-sat}
and \ref{P: 800-unsat} in Fig.~\ref{P-muB} are smooth, 
Fig.~\ref{cs2-muB} shows that the parameter set \subref{cs: 800-unsat} violates causality.
In this way, the parameter choice $(H/G, g_V/G)=(1.45, 1.2)$ in quark matter is excluded 
when $m_0=800\,\mathrm{MeV}$ in hadronic matter.

Figure~\ref{cond_check_with_Mmax} shows 
allowed combinations of $(H,g_V)$ for several choices of $m_0$.  
In all cases, the allowed values of $H$ and $g_V$ have a positive correlation;
for a larger $g_V$ we need to increase the value of $H$ \cite{Baym:2019iky}.
The details of this positive correlation depend on the low density constraint and the choice of $m_0$.
As we can see from Table.\ref{output}, the low density EOS softens for a large $m_0$,
and correspondingly smaller values of $g_V$ are favored for causal interpolations.
We note that the range of $(H,g_V)$ is larger than the previously used estimates, $(H/G,g_V/G)=(0.5, 0.5)$, based on the Fierz transformation (see e.g.~Ref. \cite{Buballa:2003qv}).
Such choices were used in the hybrid hadron-quark matter EOS with first order phase transitions, and tend to lead to the NS mass smaller than $2M_\odot$.

\section{mass-radius Relation}
\label{sec:result}

\begin{figure*}[p]
	\begin{subfigure}{\figsize}
		\includegraphics[width=\widthsize]{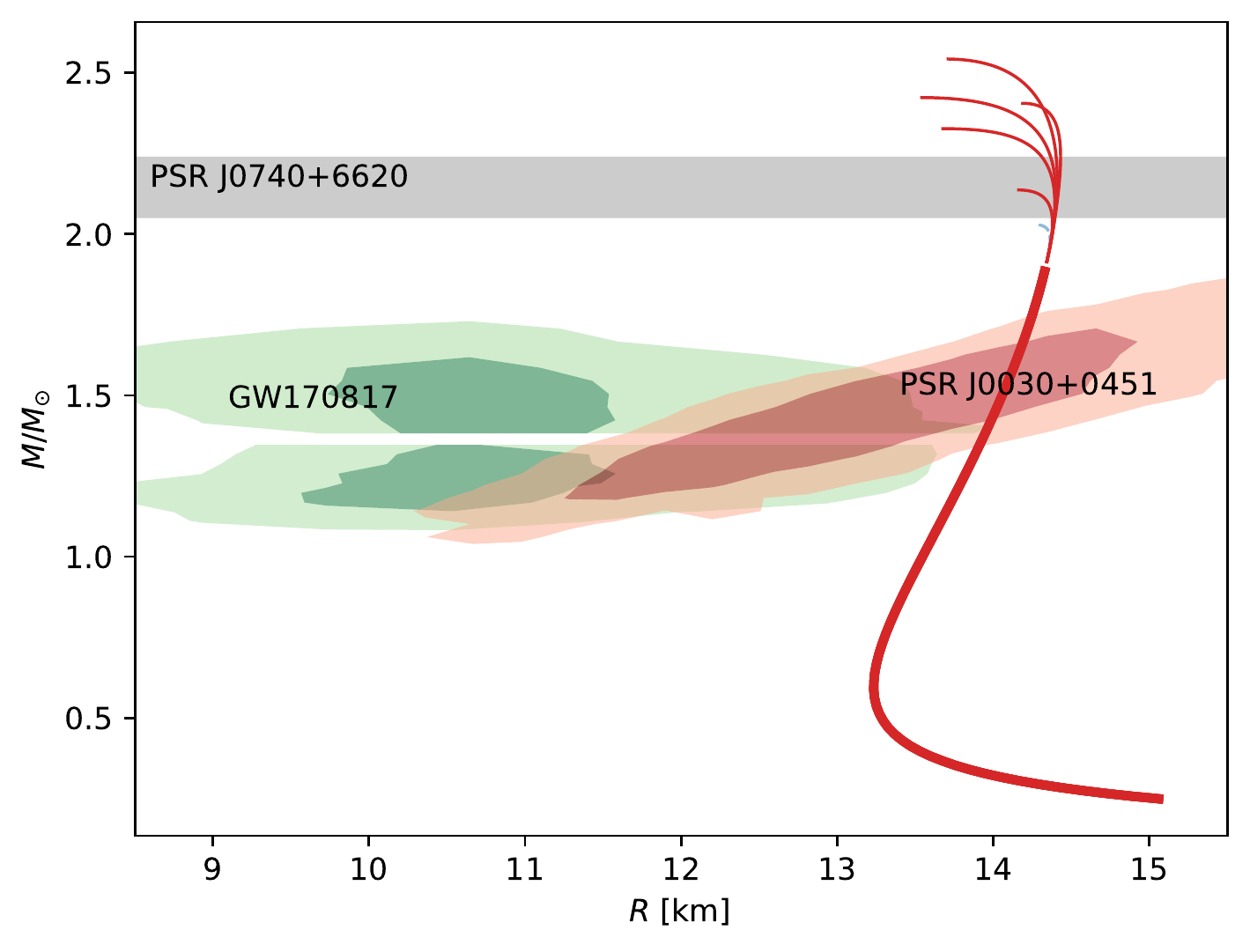}
		\caption{$m_0=500$\,MeV}
		\label{MR_500}
	\end{subfigure}
	\begin{subfigure}{\figsize}
		\includegraphics[width=\widthsize]{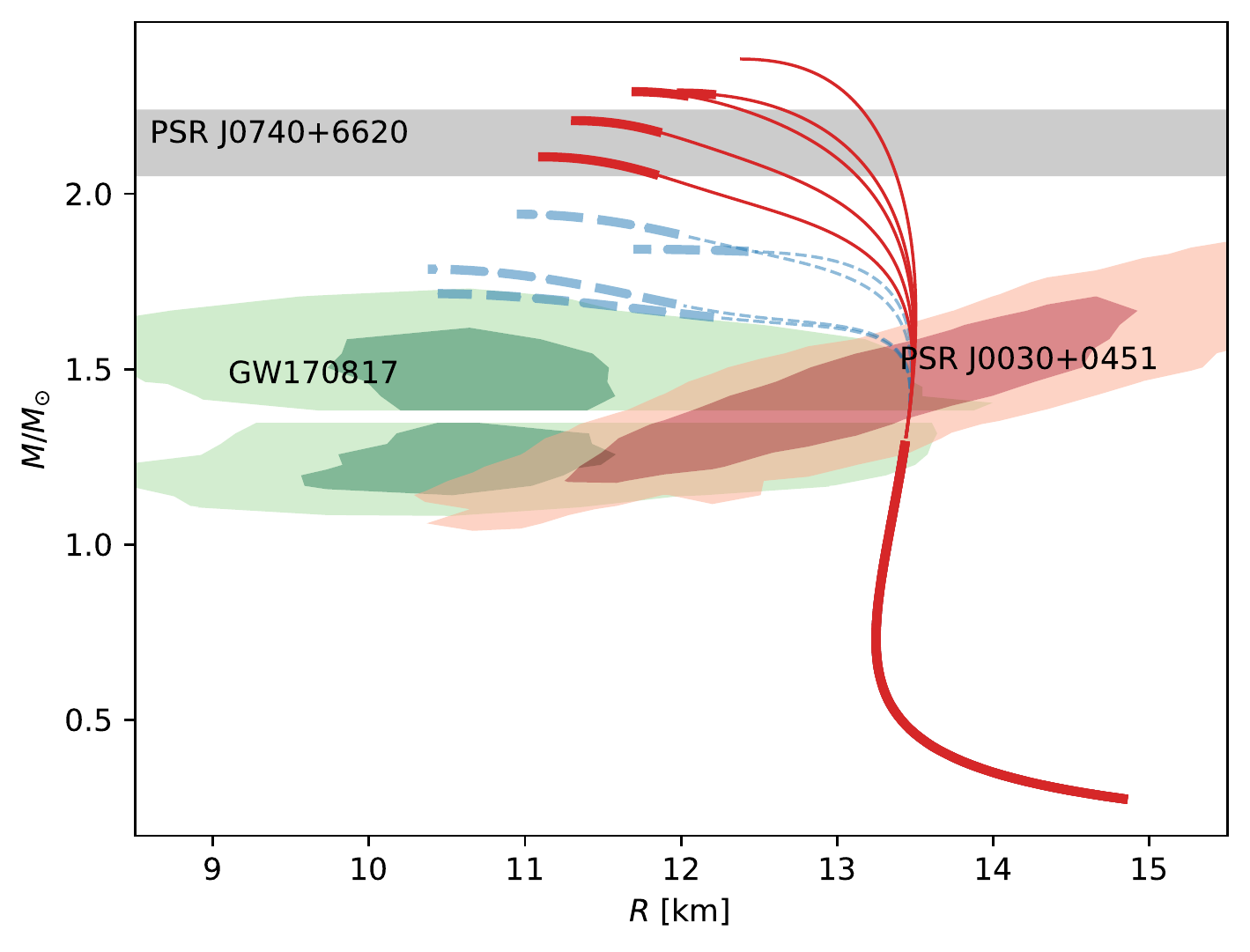}
		\caption{$m_0=600$\,MeV}
		\label{MR_600}
	\end{subfigure}
	\begin{subfigure}{\figsize}
		\includegraphics[width=\widthsize]{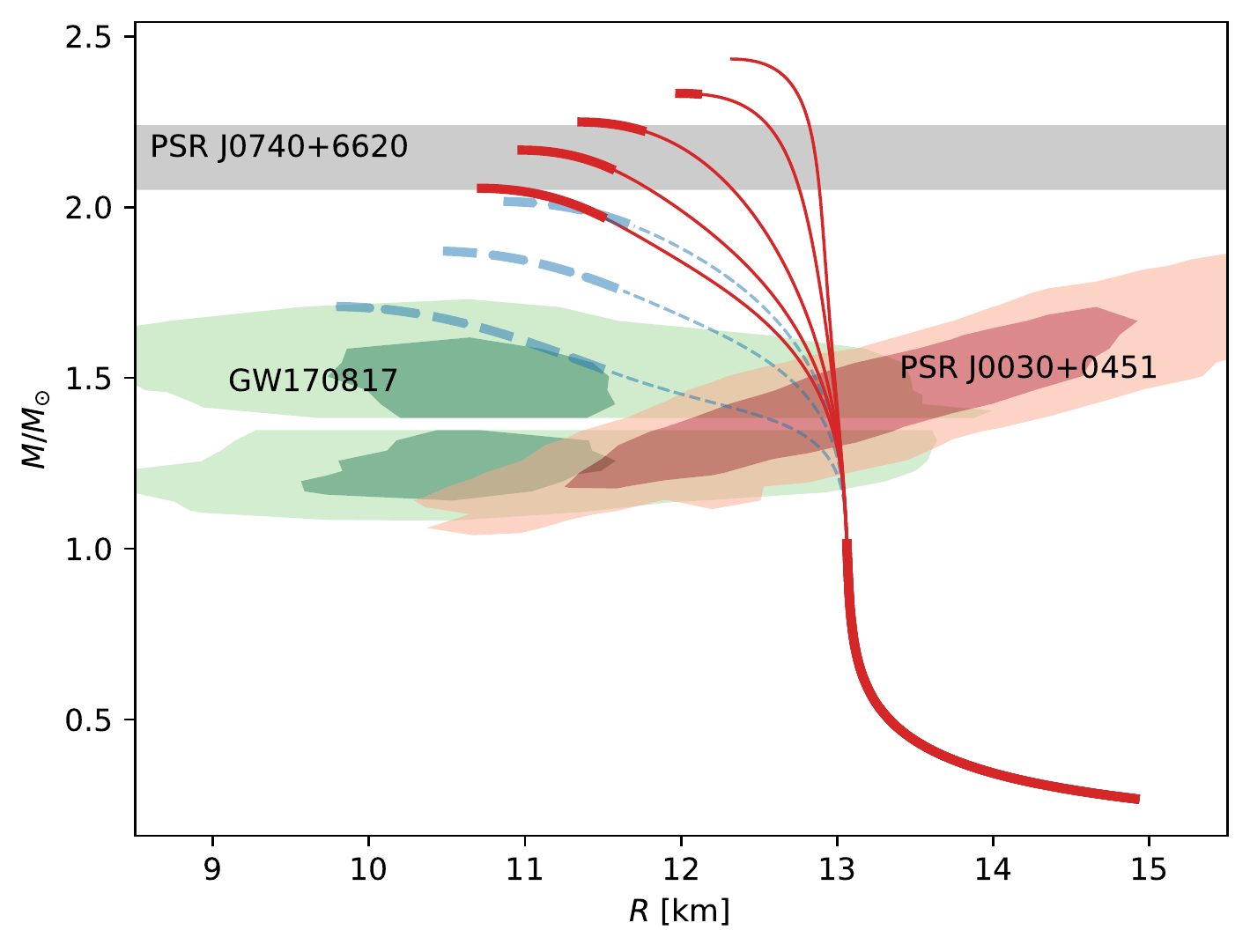}
		\caption{$m_0=700$\,MeV}
		\label{MR_700}
	\end{subfigure}
	\begin{subfigure}{\figsize}
		\includegraphics[width=\widthsize]{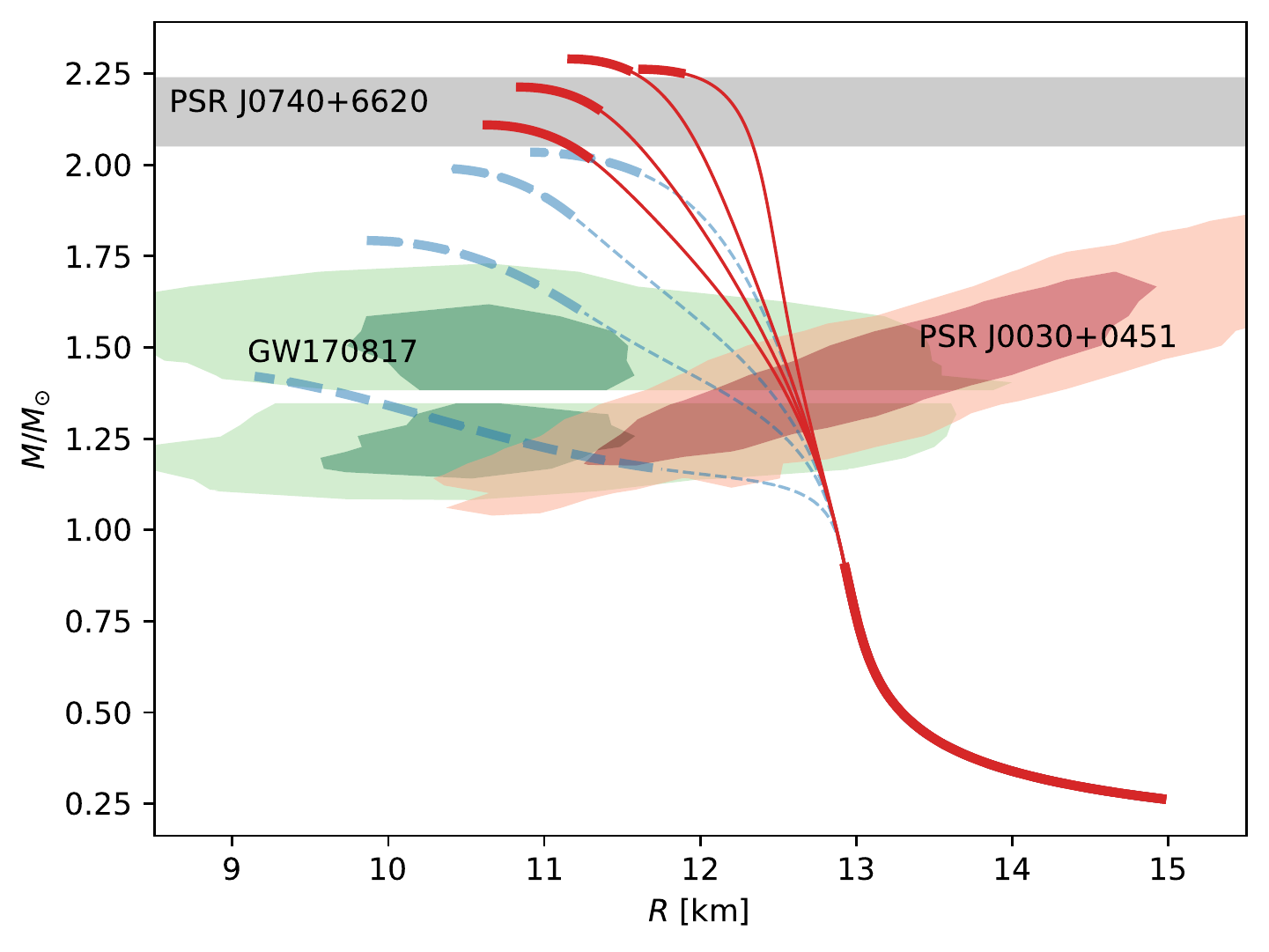}
		\caption{$m_0=800$\,MeV}
		\label{MR_800}
	\end{subfigure}
	\begin{subfigure}{\figsize}
		\includegraphics[width=\widthsize]{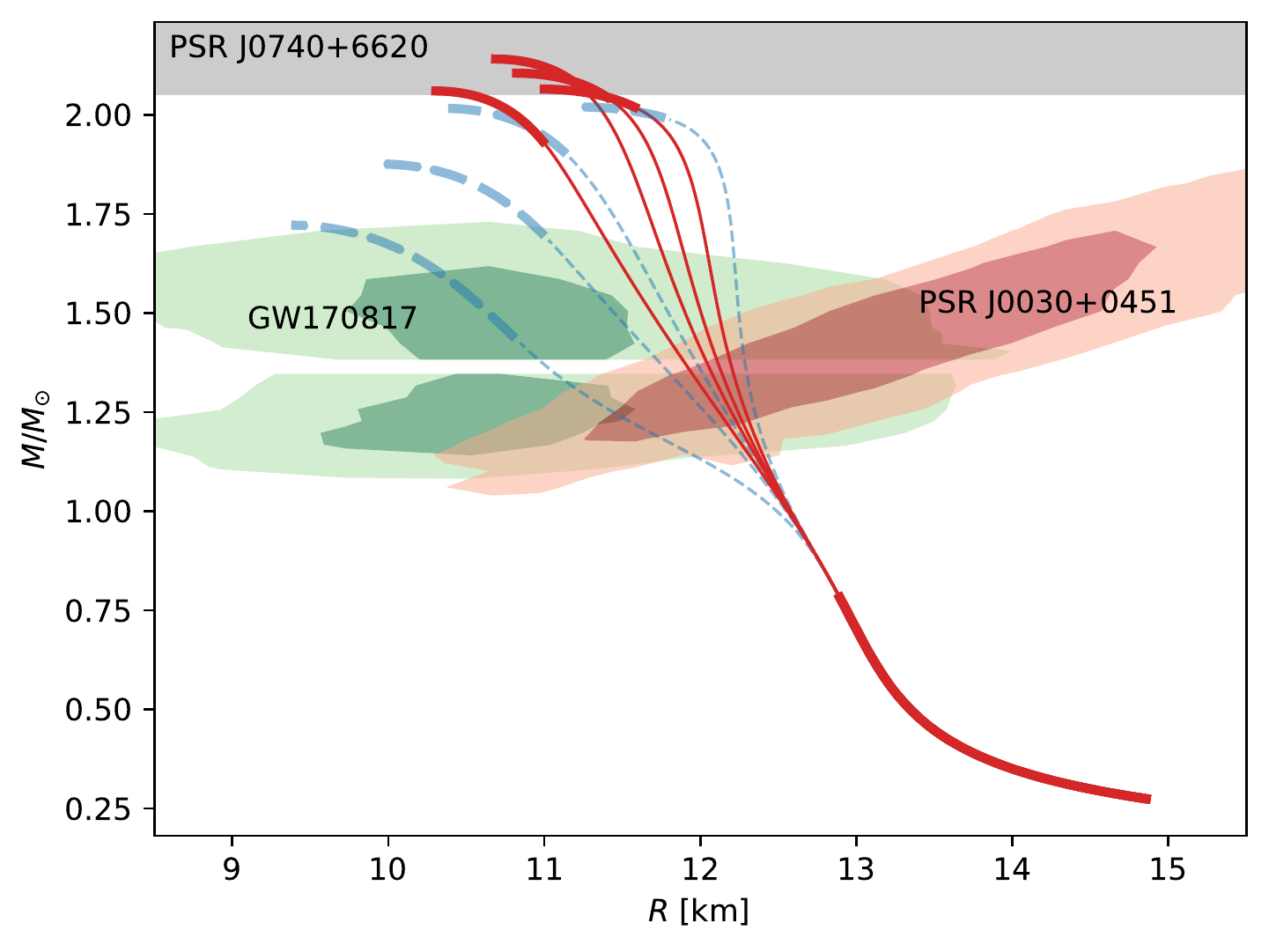}
		\caption{$m_0=900$\,MeV}
		\label{MR_900}
	\end{subfigure}
\caption{
{\small
Several choices of mass-radius relations for each $m_0$. 
(See main texts for detail.)
}
}
\label{MR}
\end{figure*}
\begin{figure*}[p]
	\begin{subfigure}{\figsize}
		\includegraphics[width=\widthsize]{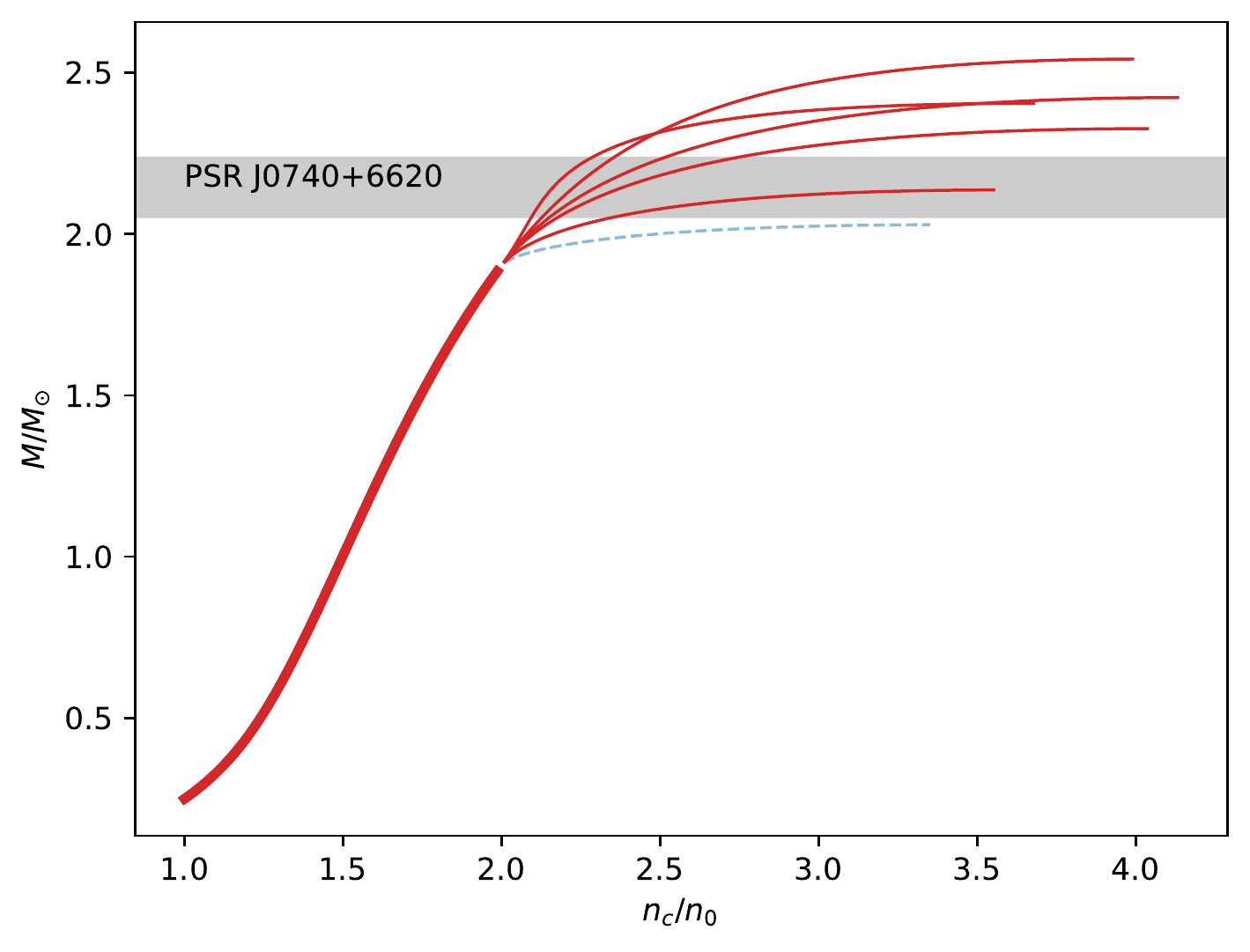}
		\caption{$m_0=500$\,MeV}
		\label{M_density_500}
	\end{subfigure}
	\begin{subfigure}{\figsize}
		\includegraphics[width=\widthsize]{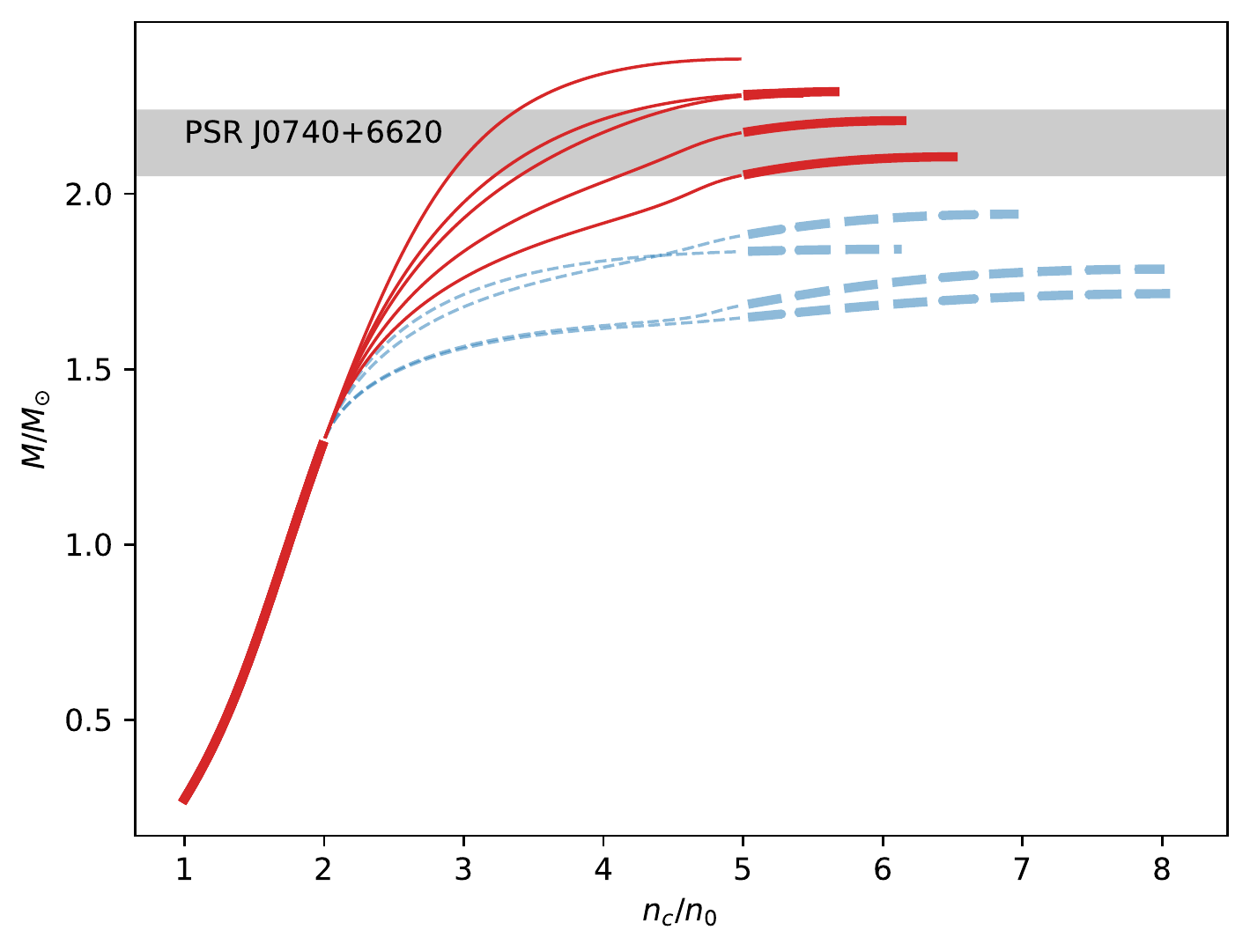}
		\caption{$m_0=600$\,MeV}
		\label{M_density_600}
	\end{subfigure}
	\begin{subfigure}{\figsize}
		\includegraphics[width=\widthsize]{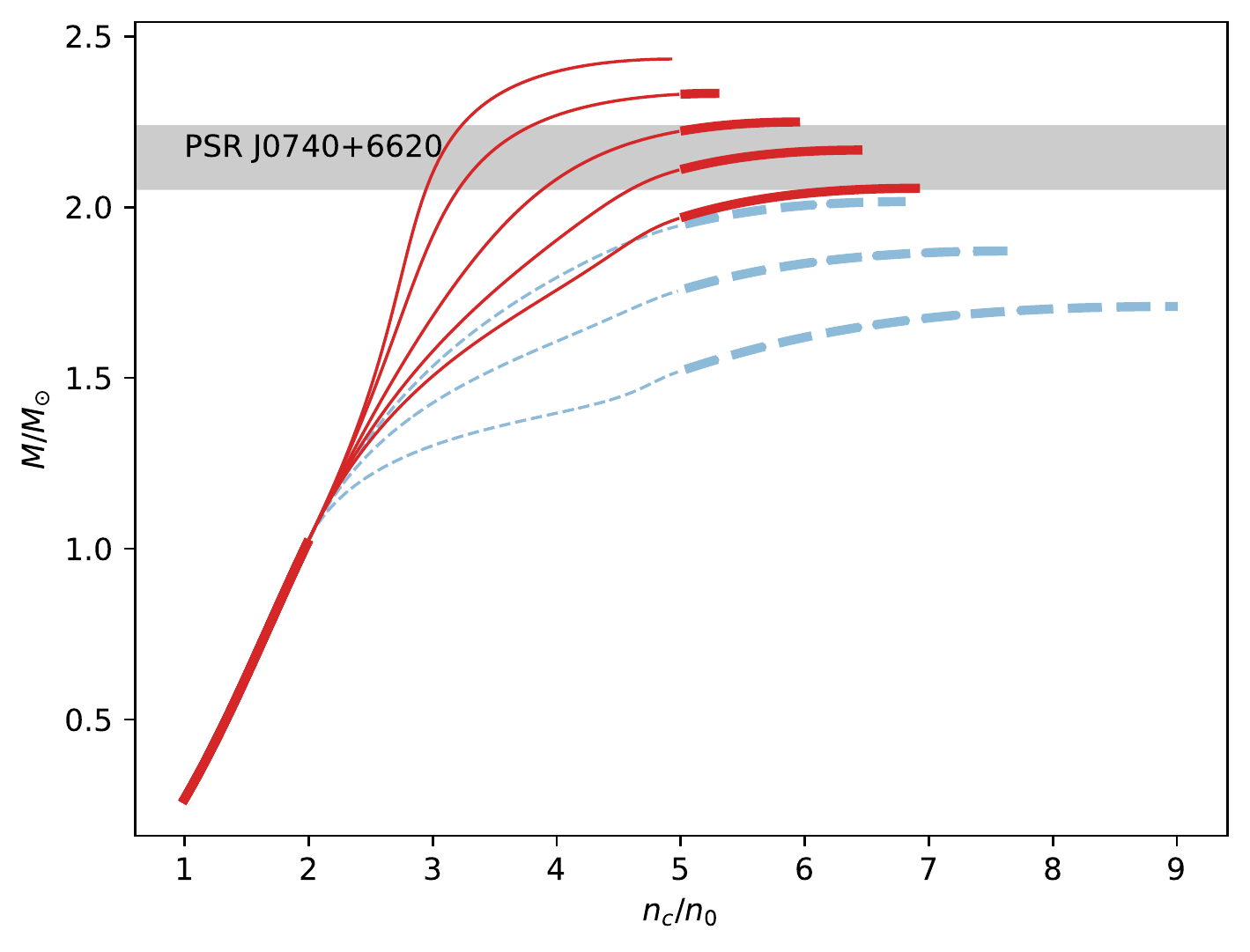}
		\caption{$m_0=700$\,MeV}
		\label{M_density_700}
	\end{subfigure}
	\begin{subfigure}{\figsize}
		\includegraphics[width=\widthsize]{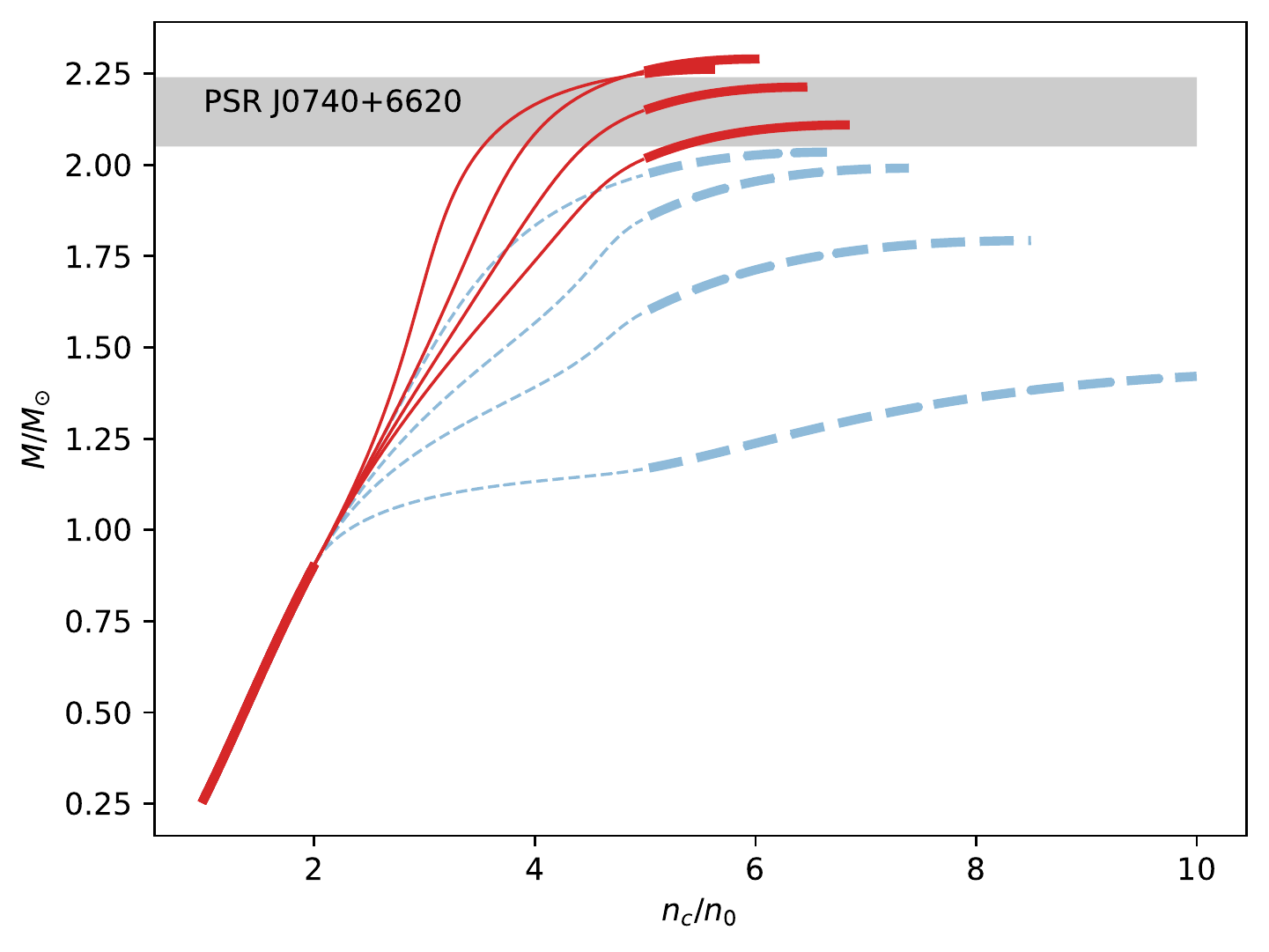}
		\caption{$m_0=800$\,MeV}
		\label{M_density_800}
	\end{subfigure}
	\begin{subfigure}{\figsize}
		\includegraphics[width=\widthsize]{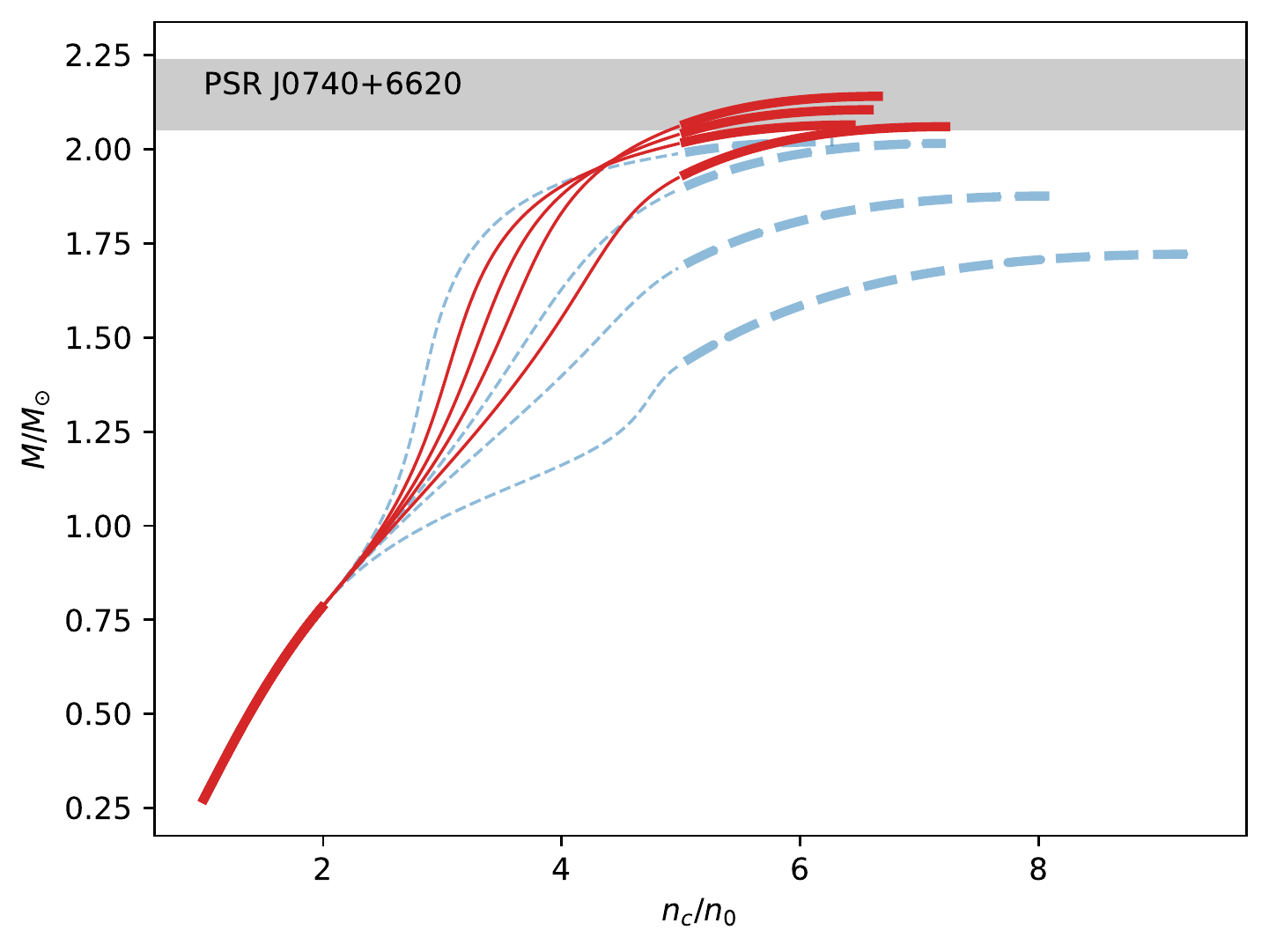}
		\caption{$m_0=900$\,MeV}
		\label{M_density_900}
	\end{subfigure}
\caption{
{\small
Several choices of relations between mass and central density for each $m_0$. 
(See main texts for detail.) 
}
}
\label{M_density}
\end{figure*}

In this section, we calculate mass-radius relations of NSs using 
the Tolman-Oppenheimer-Volkoff (TOV) equation~\cite{Tolman:1939jz,Oppenheimer:1939ne}. 
The TOV equation for hydrostatic equilibrium in general relativity is given by
\begin{align}
	\begin{aligned}
	\dv{P}{r}&=-G\frac{(\varepsilon+P)(m+4\pi r^3P)}{r^2-2Gmr} \ , \\
	\dv{m}{r}&=4\pi r^2\varepsilon \ , 
	\end{aligned}
\end{align}
where $G$ is the Newton constant, 
$r$ is the distance from the center of an NS, 
$P$, $m$ and $\varepsilon$ are the pressure, mass, and energy density 
as functions of $r$:
\begin{align}
	P=P(r) \ , \quad m=m(r) \ , \quad \varepsilon=\varepsilon(r) \ .
\end{align}
To correctly estimate NS radii, we need to include the crust equations.
We use the BPS EOS \cite{Baym:1971pw} for the outer and 
inner crust parts\footnote{
The BPS EOS is usually referred as EOS for the outer crust, but it also contains the BPP EOS \cite{Baym:1971pw} for the inner crust. 
}
at $n_B \leq 0.1\,\mathrm{fm}^{-3}$,
and at $n_B \geq 0.1\,\mathrm{fm}^{-3}$ we use our unified EOS from nuclear liquid to quark matter.

Given the central density as an initial value, 
the corresponding radius $R$ and mass $M$ of NS are obtained.
The radius is determined by the condition that the pressure vanishes: $P(R)=0$, 
and the mass is the value of $m$ at the radius: $M=m(R)$.

We show the resultant mass-radius relations in Fig.~\ref{MR}, 
and relation between mass and central density in Fig.~\ref{M_density}. 
Five panels in Figs.~\ref{MR} and \ref{M_density} correspond to five typical choices of $m_0$. 

In each panel of Figs.~\ref{MR} and \ref{M_density}, 
different curves are drawn for different combinations of $(H,g_V)$ 
indicated by circles in Fig.~\ref{cond_check_with_Mmax}. 
Thick curves in the low-mass region in Figs.~\ref{MR} and \ref{M_density}
indicate that central density of the NS is smaller than $2n_0$, 
and that the NS is made only from hadronic matter. 
Thick curves in high-mass region, on the other hand, 
imply that central density is larger than $5n_0$, 
and that core of the NS includes quark matter. 
Thin curves show that the core is in the crossover domain. 

For each combination of $(H, g_V)$, 
the maximum mass of a NS is determined, 
which are indicated by 
the color
in Fig.~\ref{cond_check_with_Mmax}. 
This shows that a larger $g_V$ or a smaller $H$ leads to a larger maximum mass. 

In this paper we use 
the mass of the millisecond pulsar PSR J0740+6620~\cite{Cromartie:2019kug}
\begin{align}\label{cond: mass}
	M_\mathrm{TOV}^\mathrm{lowest}=2.14^{+0.10}_{-0.09}\,M_\odot \ ,
\end{align}
as the lowest maximum mass, which is shown by 
gray-shaded area in Figs.~\ref{MR} and \ref{M_density}. 
Each red solid curve in these figures exhibits the mass-radius relation
for which maximum mass is larger than the above lowest maximum mass, 
while the maximum masses for mass-radius relations by blue dashed curves do not exceed the lowest maximum mass. 
We also show the constraint to the radius obtained from 
the LIGO-Virgo~\cite{TheLIGOScientific:2017qsa,GBM:2017lvd,Abbott:2018exr}
by green shaded areas on the middle 
left
\footnote{
More precisely, the LIGO-Virgo constrains the tidal deformability $\tilde{\Lambda}$ which
is the function of the tidal deformability of each NS ($\Lambda_1$ and $\Lambda_2$) and the mass ratio $q=M_2/M_1$.
But for EOS which do not lead to large variation of radii for $M \gtrsim 1M_\odot$, 
$\tilde{\Lambda}$ is insensitive to $q$. In fact the NS radii and $\tilde{\Lambda}$ can be strongly correlated
(for more details, see Ref.\cite{De:2018uhw,Radice:2017lry}),
and for our purposes it is sufficient to directly use the estimates on the radii given in Ref. \cite{Abbott:2018exr}, rather than $\tilde{\Lambda}$.
}
and from the NICER (Miller et al. \cite{Miller:2019cac} 
by red shaded areas on the middle right. 
The inner contour of each area contains $68\%$ of the posterior probability ($1\sigma$), 
and the outer one contains $95\%$ ($2\sigma$). 
These values (plus another NICER result of Riley et al. \cite{Riley:2019yda}) are summarized in Table \ref{table:radii}. 
\begin{table}[htp]
\caption{  
	{\small The radius constraints.}
}	
\begin{center}
\begin{tabular}{c|c|c |}
 &  radius [km] & mass [$M_\odot$] \\
 \hline
~~ GW170817 (primary) ~~&~~     $10.8_{-1.7}^{+2.0} $     ~~&~~  $1.46_{-0.10}^{+0.12} $ ~~ \\
~~ GW170817 (secondary) ~~&~~     $10.7_{-1.5}^{+2.1} $     ~~&~~  $1.27_{-0.09}^{+0.09} $ ~~ \\

~~ J0030+0451 (NICER \cite{Miller:2019cac}) ~~&~~ $13.02_{-1.06}^{+1.24}$ ~~&~~ $1.44_{-0.14}^{+0.15} $ ~~ \\
~~ J0030+0451 (NICER \cite{Riley:2019yda}) ~~&~~ $12.71_{-1.19}^{+1.14}$ ~~&~~ $1.34_{-0.16}^{+0.15} $ ~~ 
\end{tabular}
\end{center}
\label{table:radii}
\end{table}

In the LIGO-Virgo results which are based only on model-independent analyses, the radius of NS with $\simeq 1.4M_\odot $ is in the range of 9-13 km. If we require only our $M$-$R$ curves to be within the $2 \sigma$ band, we get the constraint $m_0 \gtrsim 600$ MeV irrespective to the quark EOS. If we further demand the $M$-$R$ curves to be within the $1 \sigma$ band, we found that only few curves with $m_0 \ge 700$ MeV meet the requirement, but those curves do not satisfy the $2M_\odot$ constraints and must be rejected.
We note that another analyses by the LIGO-Virgo suggests $11.9 \pm 1.4$ km by utilizing particular parameterization of EOS and imposing the $2M_\odot$ constraint.
 Meanwhile it is easier to reconcile our modeling with the NICER constraints which suggest larger radii, and the range $500 \le m_0 [{\rm MeV}] \le 900$ are within the $1\sigma$ band and hence do not impose further constraints in addition to the LIGO-Virgo's.
Taking into account all these results, we decided to use the $2\sigma$ band of the LIGO-Virgo results, which are compatible with available constraints, and make conservative estimates on the chiral invariant mass as
\begin{align}\label{bound m0}
	600\,\mathrm{MeV}\lesssim m_0\lesssim900\,\mathrm{MeV} \,.
\end{align}
We also note that the larger $m_0$ leads to smaller slope parameter for the symmetry energy,
$80.08 \lesssim  L_0\, [{\rm MeV}] \lesssim 86.24$, as one can read off from Table. \ref{output}.

\section{Summary and Discussions}
\label{sec:summary}

\begin{figure}[htb]
	\includegraphics[width=\widthsize]{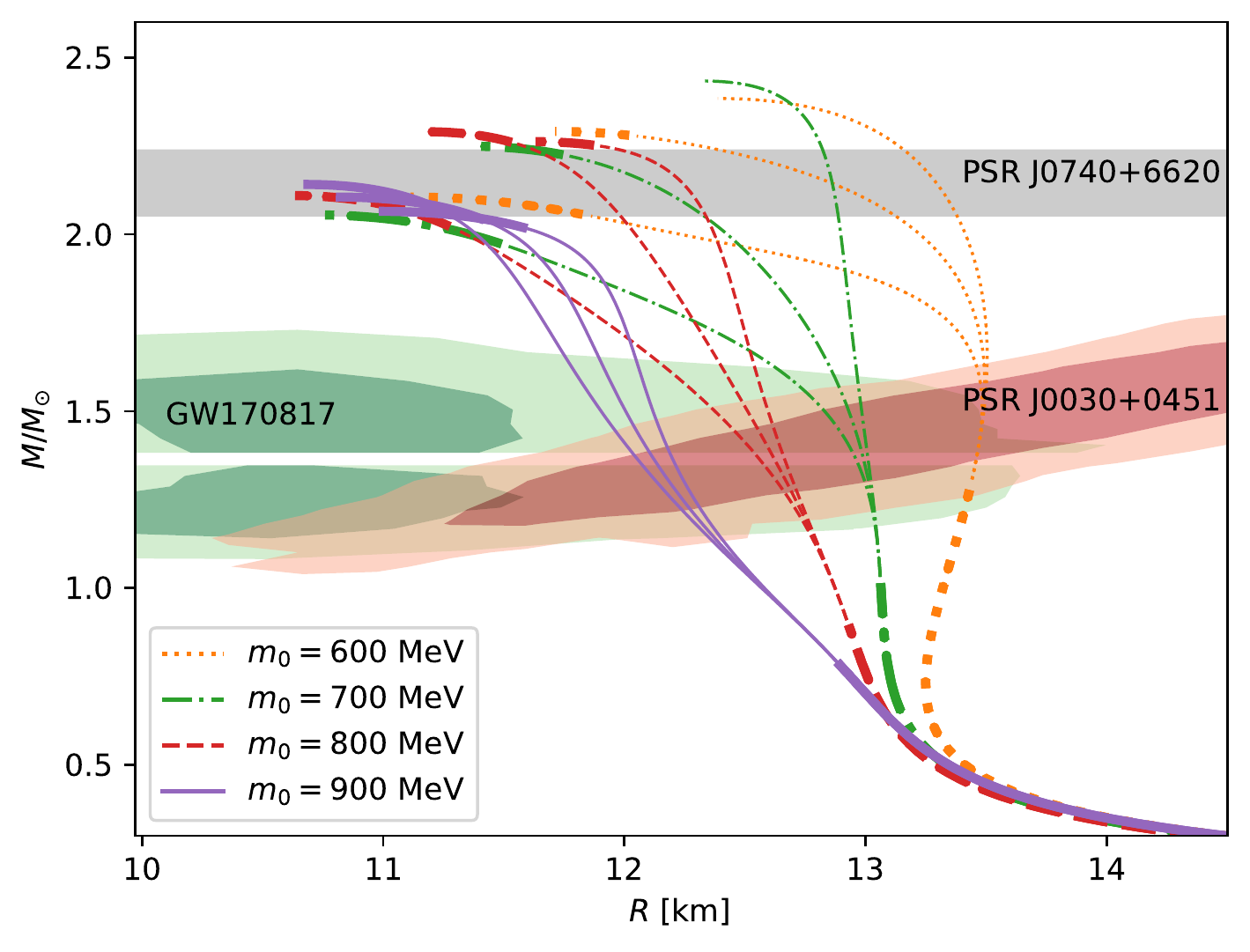}
	\caption{
{\small
Several choices of
mass-radius relations which satisfy 
both the maximum mass and the radius constraints. 
}}
\label{satisfied_all_MR}
\end{figure}

We construct EOS for NS matter by interpolating 
the EOS obtained in the PDM and the one in the NJL-type model. 
We obtain constraints to the model parameters from
thermodynamic stability, causality and the constraints on $M$-$R$ curves.

Our primary purpose was to examine how neutron star observations constrain a hadronic EOS and the microphysics in it.
Our hadronic EOS are tuned to reproduce the physics at the saturation density, but its extrapolation toward higher density is sensitive to the chiral invariant mass $m_0$.
The radii of 1.4 $M_\odot$ NS are known to have strong correlations with the stiffness of low density EOS beyond the saturation density, $n_B =1$-$2n_0$, and indeed we have obtained the nontrivial constraint,
$600 \lesssim m_0\, [{\rm MeV}] \lesssim 900$. 
At low density, the density dependence of the stiffness is sensitive to the balance between the $\sigma$- and $\omega$-exchanges, where the strength of the former strongly depends on the fraction of the chiral variant component in the nucleon mass.

Meanwhile, the maximum NS mass is known to have strong correlations with high density EOS, and constrains quark model parameters $(H, g_V)$. But these parameters are not independent of the hadronic sector, since the high and low density EOS must have a causal and thermodynamically stable connection. The allowed range of $(H, g_V)$ is sensitive to our choice of $m_0$ or the stiffness of the hadronic EOS. 
Soft hadronic EOS associated with large $m_0$ have more tensions with sufficiently stiff quark EOS, setting the upperbound $m_0 \lesssim 900 $ MeV. This upperbound is close to the total nucleon mass $m_N\approx939$ MeV, and hence is not as remarkable as the radius constraint. 

We would like to note that, as one can see in Fig.~\ref{satisfied_all_MR}, 
the cores of heavy NSs with $M\approx2 M_{\odot}$ includes quark matter 
as shown by thick curves in the heavy-mass region.
On the other hand, 
the core of $1.4M_\odot$ NS is in the crossover domain of quark and hadronic matter.
As a result, variations in the radii of $1.4 M_\odot$ NS are rather small, $\Delta R \lesssim 0.5$ km, in our crossover construction of unified EOS.

In this analysis we assumed crossover between hadronic matter and quark matter.
As we see in Fig~\ref{cond_check_with_Mmax}, 
our result showed that the coupling $H$ needs to be sufficiently large to satisfy the
causality for smooth connection, as in Ref.~\cite{Baym:2017whm}. 
Such large $H$'s ($\gtrsim 1.4G$) is consistent with the $N$-$\Delta$ splitting \cite{Song:2019qoh}, and lead to the CFL phase for $n_B \gtrsim 5n_0$. 

We note that, the previous studies as in Ref.~\cite{Baym:2017whm} primarily referred to the constraint $R \lesssim 13$ km from GW170817, but then new NICER results appear, favoring the radii $\approx13$ km. 
We may relax
the condition on low density EOS and allow stiffer EOS, which 
broadens the possibility of the first order phase transitions.
In this respect, it is interesting to explicitly implement the first order transition in the interpolated domain, 
as in Refs.~\cite{Marczenko:2019trv,Marczenko:2020jma}, while taking quark and hadronic EOS as boundary conditions.

The predicted values of the slope parameter $L_0=80-94$ MeV (shown in Table~\ref{output}) are somewhat larger than
typical estimates $L_0 =30-80$ MeV, see e.g. Refs. \cite{Kolomeitsev:2016sjl,Drischler:2020hwi}.
But there are also estimates $L_0 = (109.56 \pm 36.41)$ MeV based on recent analyses of PREXII for the neutron skin thickness \cite{Reed:2021nqk},
and we are not fully sure which estimates should be taken.
While in this study we focus on the variation of $m_0$,
the value of $L_0$ can be also adjusted by adding e.g. a term proportional to $\omega^2\rho^2$ into the hadronic part.
Such modification may slightly decrease
the lower bound and/or the upper bound of $m_0$ in Eq.~(\ref{bound m0}).
We leave such extensions of our PDM model for future studies.

\section*{Acknowledgement}
The work of M.H. is supported in part by JSPS KAKENHI Grant Number 20K03927. 
T. K. was supported by NSFC Grant No. 11875144. 
We thank C. Miller and G. Raaijmakers for providing us with numerical tables of the NICER.

\end{document}